\newif\ifOneCol

\OneColtrue

\ifOneCol

\documentclass[draftclsnofoot,twoside,onecolumn,letter,12pt]{IEEEtran}
\else
\documentclass[twocolumn,10pt]{IEEEtran}
\fi

\usepackage{cite}
\usepackage{algpseudocode}
\usepackage{graphicx}
\usepackage{amsmath}
\usepackage{amsfonts}
\usepackage{epstopdf}
\usepackage{xcolor}
\usepackage{enumerate}
\usepackage{algorithm}
\usepackage{multirow}
\usepackage{transparent}
\usepackage{subfigure}
\usepackage{float}
\usepackage{url}

\usepackage{import}

\newtheorem{theorem}{Theorem}
\newtheorem{remark}{Remark}
\newtheorem{corollary}{Corollary}
\newtheorem{lemma}{Lemma}

\newcommand{\T}{\mathrm{T}}
\newcommand{\R}{\mathrm{R}}

\let\ss= \scriptscriptstyle

\begin{document}
	\title{Heterogeneous Receptors - Based Molecule Harvesting in MC: Analysis for ISI Mitigation and Energy Efficiency}
	\author{\thanks{The conference version of this paper was submitted to IEEE GLOBECOM 2023 \cite{huang2022}. Part of this paper was presented at X. Huang's PhD dissertation \cite{xyt}.}Xinyu Huang, Yu Huang, Miaowen Wen, Nan Yang, and Robert Schober
		\vspace{-10mm}
	}
	\maketitle
	\vspace{-4mm}
\begin{abstract}
	This paper investigates a spherical transmitter (TX) with a membrane covered by heterogeneous receptors of varying sizes and arbitrary locations for molecular communication (MC), where molecules are encapsulated within vesicles and released from the TX through membrane fusion. Assuming continuous vesicle generation at the TX and a transparent receiver (RX), we calculate the molecule release rate, the fraction of absorbed molecules at the TX, and the received signal at the RX. All obtained analytical expressions are functions of all receptors' locations and sizes, and are validated by particle-based simulations. Our numerical results indicate that evenly distributed receptors on the TX membrane can absorb more molecules than randomly distributed receptors or a single receptor. Furthermore, inspired by the autoreceptor functionality in synaptic communication, we incorporate a negative feedback mechanism (NFM) at the TX, such that molecule release stops after a certain period. We then derive the fraction of molecules that can be reused for the subsequent emissions when considering both NFM and molecule harvesting. Our numerical results demonstrate that incorporating NFM can reduce inter-symbol interference (ISI) while maintaining the same peak received signal as the case without NFM. Additionally, our results show that TXs incorporating both molecule harvesting and NFM can achieve a higher energy efficiency and lower error probability than TXs employing only molecule harvesting or neither functionality.
\end{abstract}
\begin{IEEEkeywords}
	\vspace{-4mm}
	Autoreceptor, channel impulse response, energy efficiency, heterogeneous receptors, inter-symbol interference mitigation, molecular communication, molecule harvesting \vspace{-0mm}
\end{IEEEkeywords}
\section{Introduction}
In recent years, advancements in biomedical engineering, nanotechnology, and the burgeoning field of the Internet of Nano-Things have led to a growing interest in molecular communication (MC). In particular, MC is an innovative research area that investigates the use of molecules as information carriers in biological and artificial systems \cite{farsad2016comprehensive}. Owing to its unique characteristics, such as low energy consumption and potential for biocompatibility, MC is particularly well-suited for in vivo applications.

While the diffusion mechanism in MC for signal transmission does not require external energy, molecule generation and release from the TX generally consume energy. For instance, in biological systems, cells produce signaling molecules by consuming adenosine triphosphate (ATP) \cite{alberts2015essential}. Since nanomachines typically operate in resource-constrained environments, such as within living organisms, minimizing energy consumption is vital for maintaining their functionality. Furthermore, energy-efficient nanomachines can execute tasks more effectively by managing power consumption and distributing resources, ultimately enhancing overall performance. In light of this, several prior studies have proposed various methods to improve the energy efficiency of MC systems. In \cite{qiu2017bacterial}, the authors utilized bacteria as mobile relays and analyzed their information delivery energy efficiency.  The authors of \cite{cheng2020energy} examined the energy efficiency of multi-hop mobile MC systems and derived their total energy consumption and throughput. In \cite{deng2016enabling,guo2017smiet}, the authors introduced a simultaneous molecular information and energy transfer technique to reduce molecule synthesis costs, where a relay can decode the received information and generate molecules for emission using absorbed molecules through chemical reactions. While these studies stand on their own merits, they have not explored the harvesting of molecules at the TX and recycling them via biochemical reactions for subsequent emission rounds to reduce energy costs.

Molecule harvesting is a known biological mechanism in synaptic communication. One typical example is reuptake \cite{rudnick1993synapse}, which involves the re-absorption of neurotransmitters in the synaptic cleft by neurotransmitter transporters on the pre-synaptic neuron. This mechanism is essential for synaptic communication as it enables the recycling of neurotransmitters and regulates the duration of signal transmission. Inspired by this biological example, a recent study \cite{ahmadzadeh2022molecule} developed a spherical molecule harvesting TX model, where the membrane is equipped with receptors capable of reacting with information molecules. Specifically, the authors considered a homogeneous TX surface and assumed that an infinite number of receptors cover the entire TX surface or a finite number of receptors of identical sizes are uniformly distributed over the TX surface. However, the analysis in \cite{ahmadzadeh2022molecule} did not address the practical case where receptors on the TX surface can have different sizes and arbitrary locations. Notably, receptor clustering \cite{duke2009equilibrium} is a known phenomenon resulting in heterogeneous receptors on the cell membrane.

In addition to energy efficiency, error performance is another critical metric for evaluating MC systems. In particular, inter-symbol interference (ISI) is the primary cause of high error probability, especially at high data rates, and numerous studies have proposed various techniques to mitigate ISI. For instance, in \cite{noel2014improving}, enzymes are utilized to degrade information molecules, thereby mitigating ISI. However, this approach to mitigating ISI comes at the expense of a lower peak received signal at the RX. In \cite{chang2017adaptive}, the authors considered a static TX and a mobile bacterium-based RX, proposing an adaptive ISI mitigation method for changing distances and time-variant channels. The RX adaptively estimates the ISI and calculates the threshold value for making decisions. From a signal processing perspective, the authors of \cite{huang2021frequency} proposed an equalizer to counter ISI. However, none of the aforementioned studies have considered mitigating ISI by controlling the release of molecules at the TX. In \cite{tepekule2015isi}, the authors considered using two types of molecules at the TX and reducing the number of released molecules to improve energy efficiency. However, energy efficiency can only be achieved for specific bit sequence patterns, such as when all previously transmitted bits are ones. Moreover, previously proposed methods and algorithms may not be compatible with nanomachines in biological environments due to their limited computational capabilities. In fact, there are examples of biological systems that prevent excessive received signals at the RX. Autoreceptors, a type of receptor on the presynaptic neuron's membrane, can regulate neurotransmitter release by sensing the concentration of the neurotransmitter in the synaptic cleft \cite{nichols2008serotonin}. When the concentration is high, the autoreceptor activates and sends a negative feedback signal to the presynaptic neuron, inhibiting further neurotransmitter release. This helps maintain an optimal neurotransmitter level in the synaptic cleft and prevents excessive stimulation. Furthermore, this negative feedback mechanism (NFM) is beneficial for energy efficiency as it prevents excessive neurotransmitter release. Thus, given its potential advantages in ISI mitigation and energy efficiency improvement, incorporating NFM in the TX model warrants investigation.

In this paper, we consider a spherical TX, whose membrane is covered by heterogeneous receptors that may have different sizes and arbitrary locations, and we assume each receptor to be fully absorbing. We model the transportation of molecules within the TX based on \cite{huang2021membrane}, where molecules are encapsulated within vesicles and are later released by the fusion of the vesicle and the TX membrane. In particular, the vesicles are generated through various cellular processes in a dynamic and continuous manner. These processes involve the recruitment of specific coat proteins for vesicle generation, the packaging of molecules into vesicles, and the budding and scission of vesicles \cite{bonifacino2004mechanisms}. Therefore, assuming that a number of molecules already exist within the TX and are impulsively released at its center, as in  \cite{huang2021membrane}, is unrealistic. Instead, in this paper, we propose a more realistic model that involves the continuous release of vesicles within the TX, where different vesicles are released at different times due to the varying time required for their generation. To explicitly examine the impact of the considered TX model on performance, we investigate the channel impulse response (CIR) for a transparent RX, accounting for the possibility of molecule degradation during propagation from the TX to the RX. In this context, we note that the analysis in \cite{ahmadzadeh2022molecule} is not applicable when considering heterogeneous boundary conditions and vesicle-based molecule release at the TX. Furthermore, we incorporate NFM at the TX, assuming that the TX stops releasing molecules after a specific time period. We then regard the molecules that are either absorbed by the TX or not released by the TX as recyclable molecules and calculate the fraction of recyclable molecules in a single emission due to molecule harvesting and NFM. In addition, we examine the transmission of a sequence of bits from the TX to the RX and investigate the error probability of the end-to-end MC system. In summary, our major contributions are as follows:
\begin{itemize}
	\item[1)] We derive the molecule release rate from the TX membrane for the case of continuous vesicle generation within the TX. We calculate the fraction of molecules absorbed by the TX, the asymptotic fraction of absorbed molecules as time approaches infinity, and the absorption rate of molecules at the TX. On the RX side, we derive the probability of a released molecule being observed by the RX and simplify this expression for cases where receptors have the same size and are evenly distributed over the TX membrane, or when there is only a single receptor. We note that all analytical expressions are dependent on the sizes and locations of all receptors.
	\item [2)] We integrate NFM at the TX and derive the fraction of molecules that can be recycled for subsequent emissions when both molecule harvesting and NFM are considered. The accuracy of the derived analytical results is validated through particle-based simulations (PBSs). We also demonstrate that the total number of molecules absorbed by the TX remains unaffected by the vesicle generation rate, and that evenly distributed receptors on the TX membrane can capture a larger number of molecules compared to randomly distributed receptors and a single receptor, respectively.
	\item [3)] Our numerical results reveal that when incorporating NFM, the fraction of recyclable molecules increases as the time when the TX stops releasing molecules decreases. We also show that choosing an appropriate time duration for molecule release can allow the received signal to approach the peak value of the scenario without NFM, while exhibiting a fast decline after reaching the peak value.
	\item [4)] Using a single-receptor TX as an example, we evaluate the bit error rate (BER) and the fraction of recyclable molecules for three scenarios: 1) The TX employs both molecule harvesting and NFM, 2) the TX employs only molecule harvesting, and 3) the TX does not employ either of them. Our results demonstrate that the TX incorporating both molecule harvesting and NFM attains superior error performance and a larger number of recyclable molecules compared with the other two cases. Furthermore, our results reveal a trade-off between error performance and energy efficiency, which can be balanced by adjusting the parameters of NFM to meet specific application requirements.
\end{itemize}

We note that the impact of NFM is only considered in Section \ref{nf}, while the analyses in Section \ref{aor} and Section \ref{aors} do not consider NFM. The remainder of this paper is organized as follows. In Section \ref{s}, we describe the system model. In Section \ref{aor}, we derive the molecule release rate and the fraction of absorbed molecules at the TX. In Section \ref{aors}, we derive the received signal at the RX. In Section \ref{nf}, we integrate NFM at the TX and calculate the fraction of recyclable molecules. We also determine the error probability of the MC system. In Section \ref{nr}, we present and discuss our numerical results. In Section \ref{con}, we draw some conclusions.

	\section{System Model}\label{s}
	\begin{figure}[!t]
		\begin{center}
			\includegraphics[height=1.5in,width=0.48\columnwidth]{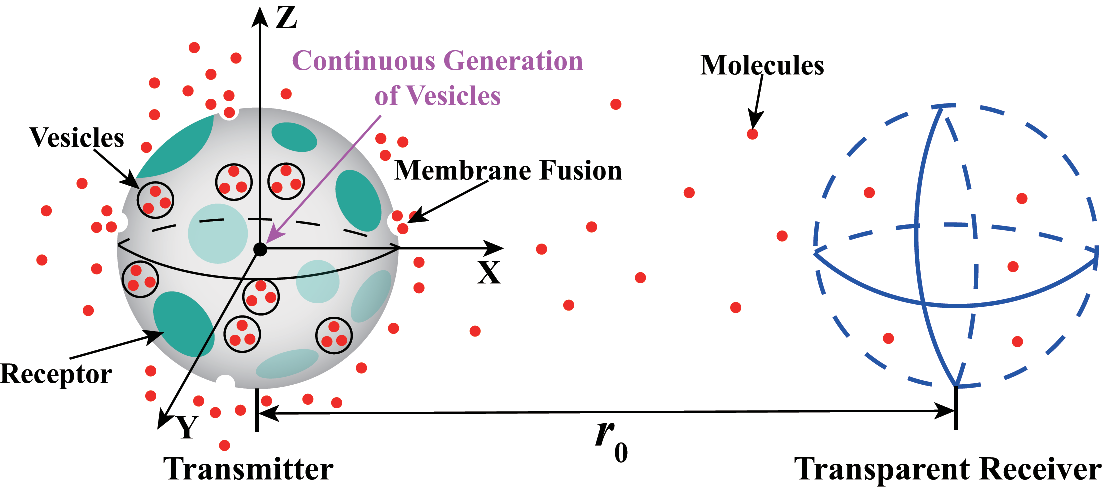}
			\caption{Illustration of the MC system model where a spherical TX covered by heterogeneous receptors communicates with a transparent RX.}\label{sys}\vspace{-0.5em}
		\end{center}\vspace{-4mm}
	\end{figure}
	In this paper, we consider an unbounded three-dimensional (3D) environment where a spherical TX communicates with a transparent spherical RX, as depicted in Fig. \ref{sys}. We choose the center of the TX as the origin of the environment and denote the radii of the TX and RX by $r_{\ss\T}$ and $r_{\ss\R}$, respectively.  
	\subsection{TX Model}\label{tm}
	In this subsection, we present the TX model in terms of vesicle generation, molecule propagation, and the receptors on the TX membrane.
	\subsubsection{Vesicle Generation}
	For the considered TX model, we assume that type-$\sigma$ molecules are stored within and transported by vesicles, where each vesicle stores $\eta$ molecules. We assume that the vesicles are continuously generated in the center of the TX. In particular, each vesicle is generated at a random time instant, and we approximate the generation process as a one-dimensional (1D) Poisson point process (PPP) \cite{barbour1988stein}. This approximation means that the time difference between two successive time instants when vesicles are generated is a random variable (RV) that follows an exponential distribution. We note that PPP has been frequently used to describe random biological processes. In particular, several previous studies, e.g., \cite{etemadi2019compound,fang2020characterization}, have modeled the generation of molecules as a 1D PPP. We further denote the total number of vesicles that the TX generates for a single transmission by $N_\mathrm{v}$ and the average number of vesicles generated per second by $\mu\;[\mathrm{vesicles}/\mathrm{s}]$.
	\subsubsection{Transportation of Molecules}\label{TM}
	We assume that the TX is filled with a fluid medium that has uniform temperature and viscosity. After vesicles are generated in the center of the TX, they diffuse randomly with a constant diffusion coefficient $D_\mathrm{v}$ until reaching the TX membrane. Then, these vesicles fuse with the membrane to release the encapsulated molecules. According to \cite{huang2021membrane}, we model the membrane fusion (MF) process between the vesicle and the TX membrane as an irreversible reaction with forward reaction rate $k_\mathrm{f}\;[\mu\mathrm{m}/\mathrm{s}]$. Thereby, if a vesicle hits the TX membrane, it fuses with the membrane with a probability of $k_\mathrm{f}\sqrt{\pi\Delta t/D_\mathrm{v}}$ during a time interval $\Delta t$ \cite{andrews2009accurate}. After MF, the molecules stored by the vesicles are instantaneously released into the propagation environment, i.e., the channel between TX and RX. In biological systems, cells maintain a balance in
 membrane length through the dynamic processes of exocytosis
 and endocytosis. Although these processes continually alter
 the membrane, the average cell membrane length remains unchanged over time. Therefore, in this paper, we assume a fixed size for the TX membrane.
	\subsubsection{Receptors on the TX Membrane}\label{r}
	We assume that there are $N_\mathrm{r}$ non-overlapping heterogeneous receptors distributed across the TX membrane, which may have different sizes and arbitrary locations. We assume the shape of the $i$th receptor to be a circle with radius $a_i$. We define $\mathcal{A}$ as the ratio of the total area of the receptors to the area of the TX surface, i.e., $\mathcal{A}=\sum_{i=1}^{N_\mathrm{r}}a_i^2/(4r_{\ss\T}^2).$ In a spherical coordinate system, we denote $\vec{l}_i=[r_{\ss\T}, \theta_i, \varphi_i]$ as the location of the center of the $i$th receptor, where $\theta_i$ and $\varphi_i$ represent the azimuthal and polar angles of the $i$th receptor, respectively. As the receptors are non-overlapping, the locations and radii of the receptors satisfy $|\vec{l}_i-\vec{l}_j|\geq a_i+a_j, \forall i, j\in\{1,2,...,N_\mathrm{r}\}$. In this paper, we assume that all receptors are fully absorbing and can only absorb the released type-$\sigma$ molecules. With this assumption, once a released diffusing type-$\sigma$ molecule hits a receptor, it is absorbed by the TX immediately. Furthermore, we assume that the TX membrane area that is not covered by receptors is perfectly reflective, which means that released diffusing type-$\sigma$ molecules are reflected back once they hit this part of the TX surface. In addition, MF is catalyzed by \textit{trans}-SNARE complexes \cite{bonifacino2004mechanisms} while molecule absorption involves transport proteins \cite{masson1999neurotransmitter}, indicating that these two processes involve distinct types of proteins and can be considered independent. 
	\subsection{Propagation Environment and RX Model}
	In this system, we consider a transparent spherical RX whose boundary does not impede the diffusion of molecules. The center of the RX is distance $r_0$ away from the center of the TX. We assume that the RX can perfectly count the number of molecules within its volume at time $t$ and use this value as the received signal. We also assume that the propagation environment between TX and RX is a fluid medium with uniform temperature and viscosity. Once molecules are released from the TX, they diffuse randomly with a constant diffusion coefficient $D_\sigma$. We further assume unimolecular degradation in the propagation environment, where type-$\sigma$ molecules can degrade to type-$\hat{\sigma}$ molecules that can neither be absorbed by the TX nor be observed by the RX, i.e., $\sigma\stackrel{k_\mathrm{d}}{\longrightarrow}\hat{\sigma}$ \cite[Ch. 9]{chang2005physical}, where $k_\mathrm{d}\;[\mathrm{s}^{-1}]$ is the degradation rate constant.
	\subsection{Negative Feedback Mechanism}\label{as}
	Autoreceptors are a type of receptor on the presynaptic neuron, which can regulate the release of neurotransmitters by sensing the concentration of the same neurotransmitter in the synaptic cleft \cite{nichols2008serotonin}. Upon detecting high concentrations, autoreceptors become activated and transmit a negative feedback signal to the presynaptic neuron, inhibiting further neurotransmitter release. Inspired by this biological mechanism, we incorporate NFM into our TX model to investigate its potential advantages for ISI mitigation and energy efficiency enhancement. Specifically, we assume that the TX can sense the surrounding molecule concentration levels and stop releasing molecules after time $\hat{t}$, even if not all of the $N_\mathrm{v}$ vesicles have been fully generated or some generated vesicles remain diffusing within the TX. The unreleased molecules can then be broken down or repackaged into vesicles for future emissions. 
\subsection{Bit Transmission}\label{bt}
	Information sent from the TX to the RX is encoded into a sequence of $Q$ binary bits, denoted by $\mathbf{b}_{1:Q}=[b_1, b_2, ..., b_Q]$, where $b_q$ is the $q$th bit, $q\in\{1,\ldots,Q\}$. We denote $T_\mathrm{b}$ as the bit interval length and assume that $b_1$ is transmitted at time $t=0$. We also assume that bits $0$ and $1$ occur with probabilities $\mathrm{Pr}(b_q=0)=P_0$ and $\mathrm{Pr}(b_q=1)=P_1=1-P_0$, where $\mathrm{Pr}(\cdot)$ denotes probability. We adopt on-off keying for modulation. Thus, $N_\mathrm{v}$ vesicles start being continuously generated from the beginning of the bit interval for bit 1, while no vesicles are generated for bit 0. We further assume that TX and RX are perfectly synchronized by some existing MC synchronization scheme, e.g., those proposed in \cite{jamali2017symbol}. To demodulate $b_q$ at the RX, we assume that the RX performs a single detection in the $q$th interval. We adopt the widely-used threshold detector that compares the number of molecules observed within the RX at the detection time with a decision threshold \cite{kuran2020survey}.

	\section{Analysis of Release and Harvest of Molecules at TX}\label{aor}
	In this section, we first analyze the release of molecules from the TX when jointly considering the continuous generation of vesicles and the MF process at the TX membrane, and derive the molecule release rate from the TX membrane. We define the molecule release rate as the probability of molecules being released during time interval $[t, t+\delta t]$ from the TX membrane, when the vesicles storing these molecules were generated in the origin at time $t=0$. Thus, we incorporate the effect of the heterogeneous receptors on the TX membrane and analyze the absorption of molecules by the TX. We further derive the fraction of molecules absorbed by the TX until time $t$ and the molecule absorption rate at time $t$.
	\subsection{Molecule Release Rate from TX Membrane}
	We define $\tau$ as the time duration needed to generate all $N_\mathrm{v}$ vesicles, which is given by $\tau=N_\mathrm{v}/\mu$. In our previous study \cite{huang2021membrane}, we assumed that vesicles are instantaneously generated in the center of the TX and analyzed the corresponding molecule release rate $f_\mathrm{r}(t)$ in \cite[Eq. (5)]{huang2021membrane}, where we assumed the same MF process as described in Section \ref{TM}. In the following theorem, based on $f_\mathrm{r}(t)$, we derive the molecule release rate, denoted by $f_\mathrm{c}(t)$, when vesicles are continuously generated in the center of the TX.
	\begin{theorem}\label{t1}
		The molecule release rate from the TX membrane at time $t$, when vesicles are continuously generated starting at time $t=0$, is given by
		\begin{align}\label{bk}
			f_\mathrm{c}(t)=\left\{\begin{array}{lr}
				f_{\mathrm{c}, 1}(t), ~~\mathrm{if}\;0<t\leq\tau,
				\\
				f_{\mathrm{c}, 2}(t), ~~\mathrm{if}\;t>\tau,
			\end{array}
			\right.
		\end{align}
		where 
		\begin{align}\label{fc}
			f_{\mathrm{c}, 1}(t)=\frac{4r_{\ss\T}^2 k_\mathrm{f}\mu}{N_\mathrm{v}D_\mathrm{v}}\sum_{n=1}^{\infty}\frac{\lambda_nj_0(\lambda_nr_{\ss\T})}{2\lambda_nr_{\ss\T}-\mathrm{sin}(2\lambda_nr_{\ss\T})}\left(1-\exp\left(-D_\mathrm{v}\lambda_n^2t\right)\right),
		\end{align}
		and
		\begin{align}\label{fc2}
			f_{\mathrm{c}, 2}(t)=&\frac{4r_{\ss\T}^2 k_\mathrm{f}\mu}{N_\mathrm{v}D_\mathrm{v}}\sum_{n=1}^{\infty}\frac{\lambda_nj_0(\lambda_nr_{\ss\T})}{2\lambda_nr_{\ss\T}-\mathrm{sin}(2\lambda_nr_{\ss\T})}\left[\exp\left(-D_\mathrm{v}\lambda_n^2(t-\tau)\right)-\exp\left(-D_\mathrm{v}\lambda_n^2t\right)\right].
		\end{align}
		In \eqref{fc} and \eqref{fc2}, $j_0(\cdot)$ is the zeroth order spherical Bessel function of the first kind \cite{olver1960bessel} and $\lambda_n$ is obtained by solving
		\begin{align}
			D_\mathrm{v}\lambda_nj_0'\left(\lambda_nr_{\ss\T}\right)=k_\mathrm{f}j_0\left(\lambda_nr_{\ss\T}\right),
		\end{align}
		with $j_0'(z)=\frac{\mathrm{d} j_0(z)}{\mathrm{d} z}$ and $n$ is a positive integer.
	\end{theorem}
	\begin{IEEEproof}
		Please see Appendix \ref{A1}.
	\end{IEEEproof}
	
	\subsection{Molecule Harvesting at TX}\label{m}
	\begin{figure}[!t]
		\begin{center}
			\includegraphics[height=1.8in,width=0.28\columnwidth]{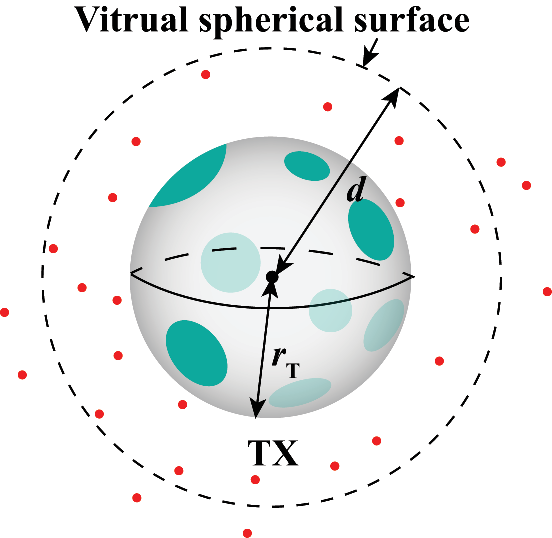}
			\caption{Molecules are uniformly released from a virtual spherical surface having distance $d$ from the center of the TX.}\label{pr2}\vspace{-0.5em}
		\end{center}\vspace{-4mm}
	\end{figure}
	In this subsection, we include the effect of heterogeneous receptors on the TX membrane, and derive the number of molecules absorbed by the TX. As vesicles are generated at the center of the TX, molecules are uniformly released from the TX membrane, i.e., the probability of molecule release is identical for any point on the TX membrane. Therefore, we first need to derive the number of molecules absorbed by the TX at time $t$ when these molecules are uniformly and simultaneously released from the TX membrane at time $t=0$, which is denoted by $H(t)$. To this end, we consider the scenario shown in Fig. \ref{pr2}, where molecules were uniformly released at time $t=0$ from the surface of a virtual sphere centered at the TX's center and having radius $d$, where $d\geq r_{\ss\T}$. The fraction of molecules absorbed by the TX in this scenario was derived in \cite[Eq. (5)]{huang2022analysis}. Based on this result, we present $H(t)$ in the following lemma.
	\begin{table}[!t]
		\newcommand{\tabincell}[2]{\begin{tabular}{@{}#1@{}}#2\end{tabular}}
		\centering
		\caption{Summary of expressions for Calculating $\frac{1}{G_{\T}}$ \cite{lindsay2017first}:}\label{tab2}
		\begin{tabular}{|c|c|}
			\hline
			\textbf{Expression}&\textbf{Size and Distribution of receptors}\\
			\hline
			\tabincell{c}{$\frac{1}{G_{\T}}=\frac{2}{N_\mathrm{r}\overline{m}\kappa r_{\ss\mathrm{T}}}\Bigg[1+\frac{\kappa}{2N_\mathrm{r}\overline{m}}
				\ln\left(\frac{\kappa}{2}\right)\sum_{i=1}^{N_\mathrm{r}}m_i^2
				+\frac{\kappa}{N_\mathrm{r}\overline{m}}\bigg(\sum_{i=1}^{N_\mathrm{r}}m_is_i$\\$+2\sum_{i=1}^{N_\mathrm{r}}
				\sum_{j=i+1}^{N_\mathrm{r}}m_im_j\mathcal{F}(\vec{l}_i',\vec{l}_j')\bigg)
				+\left(\kappa\ln\left(\frac{\kappa}{2}\right)\right)^2\frac{\vartheta}{4N_\mathrm{r}\overline{m}}
				+\mathcal{O}\left(\kappa^2\ln\left(\frac{\kappa}{2}\right)\right)\Bigg].$}& Any size, any distribution\\
			\hline
			$\frac{1}{G_\mathrm{T}}=\frac{\pi}{N_\mathrm{r}\kappa r_{\ss\mathrm{T}}}\Bigg[1+\frac{\kappa}{\pi}\bigg(\ln(2\kappa)
			-\frac{3}{2}+\frac{4}{N_\mathrm{r}}\sum_{i=1}^{N_\mathrm{r}}\sum_{j=i+1}^{N_\mathrm{r}}
			\mathcal{F}(\vec{l}_i',\vec{l}_j')\bigg)+\mathcal{O}\left(\kappa^2\ln\left(\frac{\kappa}{2}\right)\right)\Bigg]$.& Identical sizes, any distribution\\
			\hline
			$\frac{1}{G_{\T}}\approx\frac{1}{r_{\ss\T}}\left(1+\frac{\pi}{N_\mathrm{r}\kappa}+\frac{\frac{1}{2}\ln(\kappa\sqrt{N_\mathrm{r}})+\ln 2-\frac{3}{2}}{N_\mathrm{r}}-2N_\mathrm{r}^{-\frac{1}{2}}+N_\mathrm{r}^{-\frac{3}{2}}\right).$& Identical sizes, evenly distributed\\
			\hline
			$\frac{1}{G_\mathrm{T}}=\frac{\pi}{\kappa r_\mathrm{T}}\bigg[1+\frac{\kappa}{\pi}\left(\ln(2\kappa)-\frac{3}{2}\right)-\frac{\kappa^2}{\pi^2}\left(\frac{\pi^2+21}{36}\right)+\mathcal{O}(\kappa^3\ln\kappa)\bigg].$& Single receptor\\
			\hline
			
		\end{tabular}
	\end{table}  
	\begin{lemma} 
		The fraction of molecules absorbed by the TX at time $t$, when the molecules are uniformly and simultaneously released from the TX membrane at time $t=0$, is given by
		\begin{align}\label{Gt}
			H(t)=\frac{ w\mathrm{erf}(\sqrt{k_\mathrm{d}t})}{\sqrt{k_\mathrm{d}D_\sigma}}-\frac{w\gamma}{\zeta}\Big(\exp(\zeta t)\mathrm{erfc}\left(\gamma\sqrt{D_\sigma t}\right)+\gamma\sqrt{\frac{D_\sigma}{k_\mathrm{d}}}\mathrm{erf}\left(\sqrt{k_\mathrm{d}t}\right)-1\Big),
		\end{align}
		where $w=D_\sigma G_{\ss\T}/(r_{\ss\T}(r_{\ss\T}-G_{\ss\T}))$, $\gamma=1/(r_{\ss\T}-G_{\ss\T})$,  $\zeta=\gamma^2D_\sigma-k_\mathrm{d}$, $\mathrm{erf}(\cdot)$ is the error function, and $\mathrm{erfc}(\cdot)$ is the complementary error function. According to the definition and explanation in \cite{berg1977physics}, $G_{\ss\T}$ can be treated as the ``capacitance" of the TX, which is determined by the locations and sizes of the receptors. We note that $G_{\ss\T}$ measures the ability of the TX to absorb molecules given the distribution of the receptors. The expressions for $G_{\ss\T}$ for different distributions and sizes of receptors are summarized in Table \ref{tab2}, where $\kappa=\frac{a_1}{r_{\ss\mathrm{T}}}$, $m_i=\frac{2a_i}{r_{\ss\mathrm{T}}\kappa\pi}$, $\overline{m}=\frac{1}{N_\mathrm{r}}\sum_{i=1}^{N_\mathrm{r}}m_i$, $s_i=\frac{m_i}{2}\left(\ln\left(\frac{4a_i}{r_{\ss\mathrm{T}}\kappa}\right)-\frac{3}{2}\right)$, $\vartheta=\frac{\left(\sum_{i=1}^{N_\mathrm{r}}m_i^2\right)^2}{N_\mathrm{r}\overline{m}}-\sum_{i=1}^{N_\mathrm{r}}m_i^3$,
		\begin{align}\label{F}
			\mathcal{F}(\vec{l}_i', \vec{l}_j')=\left[\frac{1}{|\vec{l}_i'-\vec{l}_j'|}+\frac{1}{2}\ln|\vec{l}_i'-\vec{l}_j'|-\frac{1}{2}\ln\left(2+|\vec{l}_i'-\vec{l}_j'|\right)\right]
		\end{align}
		with $\vec{l}_i'=\vec{l}_i/r_{\ss\mathrm{T}}$, and $\mathcal{O}(\cdot)$ in Table \ref{tab2} represents the infinitesimal of higher order and is omitted during calculation. 
	\end{lemma}
	\begin{IEEEproof}
		We denote $H_\mathrm{u}(t)$ as the fraction of molecules absorbed by the TX at time $t$, when the molecules are uniformly released from a virtual spherical surface at time $t=0$, as shown in Fig. \ref{pr2}. Then, $H_\mathrm{u}(t)$ is given by \cite[Eq. (5)]{huang2022analysis}. When we set $d\rightarrow r_{\ss\T}$, the virtual spherical surface approaches to the TX membrane such that molecules can be regarded as released from the TX membrane. Therefore, $H(t)$ can be obtained as 
		\begin{align}\label{htl}
			H(t)=\lim\limits_{d\rightarrow r_{\ss\T}}H_\mathrm{u}(t).
		\end{align}
		By substituting \cite[Eq. (5)]{huang2022analysis} into \eqref{htl}, we obtain \eqref{Gt}.
	\end{IEEEproof}
	
	We then denote $H_\mathrm{e}(t)$ as the fraction of molecules absorbed by the TX at time $t$ when vesicles are continuously generated in the center of the TX starting at time $t=0$. Based on $H(t)$, we derive $H_\mathrm{e}(t)$ in the following theorem.
	\begin{theorem}\label{t2}
		The fraction of molecules absorbed at the TX by time $t$ when vesicles are continuously generated in the center of the TX starting at time $t=0$ is given by
		\begin{align}\label{ge}
			H_\mathrm{e}(t)=f_\mathrm{c}(t)*H(t),
		\end{align}
		where $*$ denotes convolution throughout this manuscript, and $f_\mathrm{c}(t)$ and $H(t)$ are given by \eqref{bk} and \eqref{Gt}, respectively.
	\end{theorem}
	\begin{IEEEproof}
		When molecules are released from the TX membrane at time $u$, $0\leq u\leq t$, the fraction of these molecules that can be absorbed by the TX is given by $f_\mathrm{c}(u)H(t-u)$. Therefore, $H_\mathrm{e}(t)$ can be expressed as $H_\mathrm{e}(t)=\int_{0}^{t}f_\mathrm{c}(u)H(t-u)\mathrm{d}u$, which can be further rewritten as the convolution in \eqref{ge}.
	\end{IEEEproof}
	
We next denote $H_{\mathrm{e}, \infty}$ as the asymptotic fraction of molecules absorbed by the TX as $t\rightarrow\infty$. We present $H_{\mathrm{e}, \infty}$ in the following corollary.
	\begin{corollary}\label{c1}
		As $t\rightarrow\infty$, the asymptotic fraction of molecules absorbed at the TX when vesicles are continuously generated in the center of the TX is given by
		\begin{align}\label{hei}
			H_{\mathrm{e}, \infty}=\frac{w}{\sqrt{D_\sigma k_\mathrm{d}}}-\frac{w\gamma^2}{\zeta}\sqrt{\frac{D_\sigma}{k_\mathrm{d}}}+\frac{w\gamma}{\zeta}.
		\end{align}
		\begin{IEEEproof}
			Please see Appendix \ref{A7}.
		\end{IEEEproof}
	\end{corollary}
	\begin{remark}\label{fe}
		From \eqref{hei}, we observe that $H_{\mathrm{e}, \infty}$ is only determined by the size of the TX, the distribution of the receptors on the TX membrane, and the diffusion coefficient and degradation rate of the molecules. It does not depend on the generation rate and diffusion coefficient of vesicles.
	\end{remark}
	
	We further denote $h_\mathrm{e}(t)$ as the molecule absorption rate at the TX membrane at time $t$, which is defined as the probability of molecules hitting receptors during the time interval $[t, t+\delta t]$. We present $h_\mathrm{e}(t)$ in the following corollary.
	\begin{corollary}
		The molecule absorption rate at the TX membrane at time $t$ when vesicles start being continuously generated in the center of the TX at time $t=0$ is given by
		\begin{align}\label{he}
			h_\mathrm{e}(t)=f_\mathrm{c,d}(t)*H(t),
		\end{align}
		where 
		\begin{align}\label{bkd}
			f_\mathrm{c,d}(t)=\left\{\begin{array}{lr}
				\frac{4r_{\ss\T}^2 k_\mathrm{f}\mu}{N_\mathrm{v}}\sum_{n=1}^{\infty}\frac{\lambda_n^3j_0(\lambda_nr_{\ss\T})}{2\lambda_nr_{\ss\T}-\mathrm{sin}(2\lambda_nr_{\ss\T})}\exp\left(-D_\mathrm{v}\lambda_n^2t\right), ~~\mathrm{if}\;0<t\leq\tau,
				\\
				\frac{4r_{\ss\T}^2 k_\mathrm{f}\mu}{N_\mathrm{v}}\sum_{n=1}^{\infty}\frac{\lambda_n^3j_0(\lambda_nr_{\ss\T})}{2\lambda_nr_{\ss\T}-\mathrm{sin}(2\lambda_nr_{\ss\T})}\left[\exp\left(-D_\mathrm{v}\lambda_n^2t\right)-\exp\left(-D_\mathrm{v}\lambda_n^2(t-\tau)\right)\right], ~~\mathrm{if}\;t>\tau,
			\end{array}
			\right.
		\end{align}
		and $H(t)$ is given in \eqref{ge}.
	\end{corollary}
	\begin{IEEEproof}
		We derive $h_\mathrm{e}(t)$ by taking the derivative of $H_\mathrm{e}(t)$ with respect to $t$, which is given by
		\begin{align}\label{hm}
			h_\mathrm{e}(t)=\frac{\partial H_\mathrm{e}(t)}{\partial t}=H(t)*\frac{\partial f_\mathrm{c}(t)}{\partial t}.
		\end{align}
		By substituting \eqref{bk} into $\partial f_\mathrm{c}(t)/\partial t$, we obtain \eqref{bkd}.
	\end{IEEEproof}
	
	\section{Analysis of Received Signal at RX}\label{aors}
	In this section, we derive analytical expressions for the received signals at the RX in two steps. In the first step, we assume that no receptors exist on the TX surface and derive the received signal at the RX. In the second step, we determine the number of molecules that no longer arrive at the RX because they were absorbed by the receptors on the TX.
	Based on these derivations, we finally obtain the signal received at the RX.
	\subsection{Problem Formulation}\label{pf}
	\begin{figure}[!t]
		\begin{center}
			\includegraphics[height=1.2in,width=0.42\columnwidth]{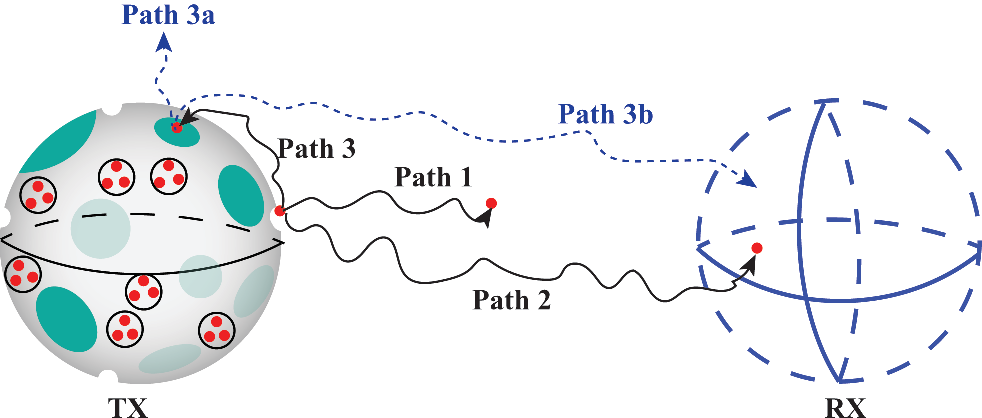}
			\caption{All possible diffusion paths of released molecules.}\label{sys2}\vspace{-0.5em}
		\end{center}\vspace{-4mm}
	\end{figure}
In this subsection, we consider two types of TXs: TX-A and TX-B. TX-A refers to the TX without receptor coverage, causing molecules to be reflected back upon collision with any point on TX-A, and TX-B represents the TX covered by receptors, as depicted in Section \ref{tm}. We define $P_{\ss\T}(t)$ and $P(t)$ as the probabilities that a released molecule is observed at the RX at time $t$ for TX-A and TX-B, respectively. Due to the presence of absorbing receptors on the TX-B, $P(t)<P_{\ss\T}(t)$, and our objective in this section is to calculate $P(t)$. Since the geometry of TX-A is simplistic, determining $P_{\ss\T}(t)$ is straightforward. By defining $P_
\mathrm{r}(t)=P_{\ss\T}(t)-P(t)$, we can calculate $P(t)$ as
\begin{align}\label{pt}
	P(t)=P_{\ss\T}(t)-P_\mathrm{r}(t),
\end{align} 
where $P_\mathrm{r}(t)$ represents the probability of fewer observed molecules at the RX due to the presence of receptors on the TX. To derive $P_\mathrm{r}(t)$, we classify all possible diffusion paths of molecules after their release from the TX membrane into three categories, namely path 1, path 2, and path 3, as shown in Fig. \ref{sys2}. Path 1 is the path where molecules diffuse in the propagation environment at time $t$, path 2 is the path where molecules move into the RX at time $t$, and path 3 is the path where molecules hit a receptor on the TX-B's surface at time $t$. For TX-A, we further divide path 3 into path 3a and path 3b. Path 3a is the path where molecules diffuse in the propagation environment at time $t$ after having hit the TX at time $u$, where $u\leq t$, and path 3b is the path where molecules move into the RX at time $t$ after having hit the TX at time $u$. However, in the case of TX-B, as the molecules get absorbed by receptors along path 3, there are no molecules that can be observed by the RX in path 3b. Therefore, $P_\mathrm{r}(t)$ represents the probability that a released molecule is observed at the RX at time $t$ when molecules move along path 3b.
	\subsection{Derivation of Received Signal}
	We first consider TX-A. Similar to the procedure in Section \ref{m}, to derive $P_{\ss\T}(t)$, we need to first derive the received signal at the RX when molecules are uniformly and simultaneously released from the TX membrane at time $t=0$, denoted by $P_\mathrm{u}(t)$. When a molecule is released from an arbitrary point $\alpha$ on the membrane of a spherical TX, the probability that this molecule is observed at the RX, denoted by $P_\alpha(t)$, is given by \cite[Eq. (27)]{noel2013using}
	\begin{align}\label{pit}
		P_\alpha(t)=&\frac{1}{2}\left[\mathrm{erf}\left(\frac{r_{\ss\R}-r_\alpha}{\sqrt{4D_\sigma t}}\right)+\mathrm{erf}\left(\frac{r_{\ss\R}+r_\alpha}{\sqrt{4D_\sigma t}}\right)\right]\exp(-k_\mathrm{d}t)+\frac{1}{r_\alpha}\sqrt{\frac{D_\sigma t}{\pi}}\left[\exp\left(-\frac{(r_{\ss\R}+r_\alpha)^2}{4D_\sigma t}-k_\mathrm{d}t\right)\right.\notag\\&\left.-\exp\left(-\frac{(r_{\ss\R}-r_\alpha)^2}{4D_\sigma t}-k_\mathrm{d}t\right)\right],
	\end{align}
	where $r_\alpha$ is the distance between point $\alpha$ and the center of the RX. By taking the surface integral of $P_\alpha(t)$ over the TX membrane, we obtain $P_\mathrm{u}(t)$ presented in the following lemma.
	\begin{lemma}\label{l2}
		If there are no receptors on the TX membrane, the probability that a molecule is observed at the RX at time $t$, when these molecules were uniformly and simultaneously released from the TX membrane at time $t=0$, is given by
		\begin{align}\label{pu}
			P_\mathrm{u}(t)=&\frac{1}{8r_0r_{\ss\T}}\left[\xi_1(r_0-r_{\ss\T}, t)+\xi_1(r_{\ss\T}-r_0, t)-\xi_1(r_0+r_{\ss\T}, t)-\xi_1(-r_0-r_{\ss\T}, t)\right]+\frac{D_\sigma t}{2r_{\ss\T}r_0}\notag\\&\times\left[\xi_2(r_{\ss\T}+r_0, t)+\xi_2(-r_{\ss\T}-r_0, t)-\xi_2(r_0-r_{\ss\T}, t)-\xi_2(r_{\ss\T}-r_0, t)\right],
		\end{align}
		where 
		\begin{align}
			\xi_1(z,t)=\exp\left(-\frac{(r_{\ss\R}-z)^2}{4D_\sigma t}-k_\mathrm{d}t\right)\left(r_{\ss\R}+z\right)\sqrt{\frac{4D_\sigma t}{\pi}}+\left(r_{\ss\R}^2+2D_\sigma t-z^2\right)\mathrm{erf}\left(\frac{r_{\ss\R}-z}{\sqrt{4D_\sigma t}}\right)\exp\left(-k_\mathrm{d}t\right),
		\end{align}	
		and $\xi_2(z,t)=\mathrm{erf}\left(\frac{r_{\ss\R}+z}{\sqrt{4D_\sigma t}}\right)\exp\left(-k_\mathrm{d}t\right)$.
	\end{lemma}
	\begin{IEEEproof}
		Please see Appendix \ref{A3}.
	\end{IEEEproof}
	
Based on $P_\mathrm{u}(t)$, we present $P_{\ss\T}(t)$ in the following lemma.
	
	\begin{lemma}
		If there are no receptors on the TX membrane, the probability that a released molecule is observed at the RX at time $t$, when the vesicles are continuously generated in the center of the TX starting from time $t=0$, is given by
		\begin{align}\label{pt1}
			P_{\ss\T}(t)=f_\mathrm{c}(t)*P_\mathrm{u}(t),
		\end{align} 
		where $f_\mathrm{c}(t)$	and $P_\mathrm{u}(t)$ are given in \eqref{bk} and \eqref{pu}, respectively. 
		\begin{IEEEproof}
			The proof is similar to the proof of Theorem \ref{t2}, and thus omitted here.
		\end{IEEEproof}
	\end{lemma}
	
	Next, we derive the probability that a released molecule is observed at the RX associated with path 3b. As we assume that $\mathcal{A}$ and $a_i$ are extremely small compared to the distance between TX and RX, each receptor can be regarded as a point TX which releases molecules with a release rate that equals the hitting rate of molecules on this receptor at time $t$. As molecules are uniformly released from the TX membrane, the probability of molecules hitting is the same for any point on the receptors. We denote $h_{\mathrm{e},i}(t)$ as the hitting rate of molecules on the $i$th receptor at time $t$, and express $h_{\mathrm{e},i}(t)$ as
	\begin{align}\label{heit}
		h_{\mathrm{e},i}(t)=\frac{\mathcal{A}_i}{\mathcal{A}}h_\mathrm{e}(t),
	\end{align}
	where $\mathcal{A}_i=a_i^2/(4r_{\ss\T}^2)$ is the ratio of the area of the $i$th receptor to the area of the TX surface, and $h_\mathrm{e}(t)$ is the total hitting rate of molecules on the TX membrane which is given by \eqref{he}. We recall that $\mathcal{A}$ is the ratio of the total area of receptors to the area of the TX surface, as mentioned in Section \ref{r}. Then, we are ready to derive $P_\mathrm{r}(t)$ and $P(t)$. We present $P_\mathrm{r}(t)$ in the following lemma.
	\begin{lemma}\label{l4}
		The probability that a released molecule is observed at the RX associated with path 3d at time $t$ is given by
		\begin{align}\label{pr}
			P_\mathrm{r}(t)=\frac{f_\mathrm{c,d}(t)*H(t)}{\mathcal{A}}*\sum_{i=1}^{N_\mathrm{r}}\mathcal{A}_iP_\alpha(t)\big|_{r_\alpha=d_i},
		\end{align}
		where $d_i$ represents the distance between the centers of the $i$th receptor and the RX, i.e., $d_i=\sqrt{r_{\ss\T}^2-2r_0r_{\ss\T}\cos(\varphi_i)\sin(\theta_i)+r_0^2}$, and $f_\mathrm{c,d}(t)$, $H(t)$, and $P_\alpha(t)$ are given by \eqref{bkd}, \eqref{Gt}, and \eqref{pit}, respectively.
	\end{lemma}
	\begin{IEEEproof}
		Please see Appendix \ref{A4}.
	\end{IEEEproof}
	
	We now have $P_{\ss\T}(t)$ and $P_\mathrm{r}(t)$. According to \eqref{pt}, we present $P(t)$ in the following theorem.
	\begin{theorem}\label{t3}
		The probability that a released molecule is observed at the RX at time $t$, when vesicles are continuously generated in the center of the TX starting from time $t=0$, is given by
		\begin{align}\label{ptf}
			P(t)=f_\mathrm{c}(t)*P_\mathrm{u}(t)-\frac{f_\mathrm{c,d}(t)*H(t)}{\mathcal{A}}*\sum_{i=1}^{N_\mathrm{r}}\mathcal{A}_iP_\alpha(t)\big|_{r_\alpha=d_i},
		\end{align}
		where $f_\mathrm{c}(t)$, $P_\mathrm{u}(t)$, $f_\mathrm{c,d}(t)$, $H(t)$, and $P_\alpha(t)$ are given in \eqref{bk}, \eqref{pu}, \eqref{bkd}, \eqref{Gt}, and \eqref{pit}, respectively.
	\end{theorem}
	
	We note that the number of observed molecules within the RX at time $t$ is $N_\mathrm{v}\eta P(t)$. For certain distributions of receptors, we can further simplify \eqref{ptf} in the following corollary.
	\begin{corollary}\label{c6}
		When all receptors have the same size and are evenly distributed over the TX membrane, \eqref{ptf} can be simplified as follows
		\begin{align}\label{pts}
			P(t)=(f_\mathrm{c}(t)-f_\mathrm{c,d}(t)*H(t))*P_\mathrm{u}(t).
		\end{align}
		When there is only a single receptor on the TX membrane, \eqref{ptf} can be simplified as
		\begin{align}\label{ptfc}
			P(t)=f_\mathrm{c}(t)*P_\mathrm{u}(t)-f_\mathrm{c,d}(t)*H(t)*P_\alpha(t)\big|_{r_\alpha=d_\mathrm{s}},
		\end{align}
		where $d_\mathrm{s}$ represents the distance between the centers of the single receptor and the RX.
	\end{corollary}
	\begin{IEEEproof}
		Please see Appendix \ref{A5}.
	\end{IEEEproof}
	
	\section{Performance Analysis under NFM and Average Error Probability}\label{nf}
	In this section, we integrate NFM at the TX and examine the resulting error performance of the system. First, we derive a closed-form expression for the fraction of recyclable molecules of a single transmission, taking into account both NFM and the molecules absorbed by the TX. Second, we assume a sequence of bits is transmitted from the TX to the RX and derive the corresponding bit error rate (BER).
	\subsection{Analysis of NFM}
	In this section, we investigate the fraction of molecules that can be reused for the subsequent molecule release when accounting for both NFM and molecule harvesting at the TX. Due to the impact of NFM, the TX stops releasing molecules after time $\hat{t}$. We then denote $\hat{f}_\mathrm{c}(t)$ as the molecule release rate when incorporating NFM and express it as
	\begin{align}\label{tbk}
		\hat{f}_\mathrm{c}(t)=\left\{\begin{array}{lr}
			\hat{f}_{\mathrm{c}, 1}(t), ~~\mathrm{if}\;\hat{t}\leq\tau,
			\\
			\hat{f}_{\mathrm{c}, 2}(t), ~~\mathrm{if}\;\hat{t}>\tau,
		\end{array}
		\right.
	\end{align}
where
\begin{align}
	\hat{f}_{\mathrm{c}, 1}(t)=\left\{\begin{array}{lr}
		f_{\mathrm{c}, 1}(t), ~~\mathrm{if}\;0\leq t\leq\hat{t},
		\\
		0, ~~~~~~~~\mathrm{if}\;\hat{t}\leq t,
		\end{array}
	\right.
\end{align}	
and
\begin{align}\label{hf}
	\hat{f}_{\mathrm{c}, 2}(t)=\left\{\begin{array}{lr}
		f_{\mathrm{c}, 1}(t), ~~\mathrm{if}\;0\leq t\leq\tau,
		\\
		f_{\mathrm{c}, 2}(t), ~~\mathrm{if}\;\tau<t<\hat{t},\\
		0, ~~~~~~~~\mathrm{if}\;t>\hat{t}.
	\end{array}
\right.
\end{align}
Furthermore, we denote $\chi(\hat{t})$ as the fraction of recyclable molecules in a single emission. Here, the pool of recyclable molecules comprises both the molecules that remain unreleased by the TX due to NFM and the molecules that are absorbed back by the TX.
 We denote $\beta_1(t)$ as the fraction of molecules absorbed by the TX at time $t$. Similar to \eqref{ge}, we express $\beta_1(t)$ as $\beta_1(t)=\hat{f}_\mathrm{c}(t)*H(t)$. We further denote $\beta_2(\hat{t})$ as the fraction of molecules that remain unreleased by the TX due to NFM. Then, we can express $\chi(\hat{t})$ as 
	\begin{align}\label{ch}
	\chi(\hat{t})=\beta_1(t\rightarrow\infty)+\beta_2(\hat{t}).
\end{align}
We derive and present $\chi(\hat{t})$ in the following theorem.
\begin{theorem}\label{t4}
	For molecule harvesting and NFM at the TX, the fraction of molecules that can be reused for the subsequent emissions is given by
	\begin{align}
		\chi(\hat{t})=\left\{\begin{array}{lr}
			\chi_1(\hat{t}), ~~\mathrm{if}\;\hat{t}\leq\tau,
			\\
			\chi_2(\hat{t}), ~~\mathrm{if}\;\hat{t}>\tau,
		\end{array}
		\right.
	\end{align}
where
\begin{align}\label{ha}
	\chi_1(\hat{t})=&\frac{4r_{\ss\T}^2 k_\mathrm{f}\mu\eta}{ D_\mathrm{v}}\sum_{n=1}^{\infty}\frac{\lambda_nj_0(\lambda_nr_{\ss\T})}{2\lambda_nr_{\ss\T}-\mathrm{sin}(2\lambda_nr_{\ss\T})}\left[\left(\frac{w}{\sqrt{D_\sigma k_\mathrm{d}}}-\frac{w\gamma^2}{\zeta}\sqrt{\frac{D_\sigma}{k_\mathrm{d}}}+\frac{w\gamma}{\zeta}-1\right)\right.\notag\\&\left.\times\left(\hat{t}-\frac{1-\exp(-D_\mathrm{v}\lambda_n^2\hat{t})}{D_\mathrm{v}\lambda_n^2}\right)+\tau\right],
\end{align}
and
	\begin{align}\label{chi}
		\chi_2(\hat{t})=&\frac{4r_{\ss\T}^2 k_\mathrm{f}\mu\eta}{ D_\mathrm{v}}\sum_{n=1}^{\infty}\frac{\lambda_nj_0(\lambda_nr_{\ss\T})}{2\lambda_nr_{\ss\T}-\mathrm{sin}(2\lambda_nr_{\ss\T})}\left[\frac{w\tau}{\sqrt{D_\sigma k_\mathrm{d}}}-\frac{w\gamma^2\tau}{\zeta}\sqrt{\frac{D_\sigma}{k_\mathrm{d}}}+\frac{w\gamma\tau}{\zeta}\right.\notag\\&\left.+\frac{\exp\left(-D_\mathrm{v}\lambda_n^2(\hat{t}-\tau)\right)-\exp\left(-D_\mathrm{v}\lambda_n^2\hat{t}\right)}{D_\mathrm{v}\lambda_n^2}\left(1-\frac{w}{\sqrt{D_\sigma k_\mathrm{d}}}+\frac{w\gamma^2}{\zeta}\sqrt{\frac{D_\sigma}{k_\mathrm{d}}}-\frac{w\gamma}{\zeta}\right)\right].
	\end{align}
\end{theorem}
\begin{IEEEproof}
	Please see Appendix \ref{A6}.
\end{IEEEproof}
\subsection{Analysis of Error Performance}
In this subsection, we assume a sequence of bits is transmitted from the TX to the RX, as described in Section \ref{bt}, and analyze the detection performance at the RX. We denote $\hat{P}(t)$ as the received signal at the RX at time $t$ when incorporating NFM at the TX. Similar to \eqref{ptf}, we derive $\hat{P}(t)$ as 
\begin{align}\label{tp}
	\hat{P}(t)=\hat{f}_\mathrm{c}(t)*P_\mathrm{u}(t)-\frac{\hat{h}_\mathrm{e}(t)}{\mathcal{A}}*\sum_{i=1}^{N_\mathrm{r}}\mathcal{A}_iP_\alpha(t)\big|_{r_\alpha=d_i},
\end{align}
where $\hat{h}_\mathrm{e}(t)$ is the absorption rate of molecules at the TX with NFM at time $t$. Due to the discontinuity of $\hat{f}_\mathrm{c}(t)$ in \eqref{tbk}, it is hard to derive $\hat{h}_\mathrm{e}(t)$ directly by taking the derivative of $\beta_1(t)$ with respect to $t$ as in \eqref{hm}. Here, we use the central difference method \cite{fornberg1988generation} to calculate $\hat{h}_\mathrm{e}(t)$, yielding $\hat{h}_\mathrm{e}(t)=\left(\beta_1(t+\epsilon)-\beta_1(t-\epsilon)\right)/(2\epsilon)$ with  $\epsilon=10^{-3}$. Without loss of generality, we use $\hat{P}(t)$ as an example in the following error performance analysis. For a TX without NFM, we can obtain the error performance by replacing  $\hat{P}(t)$ with $P(t)$ in the following analysis.

We denote $t_{\mathrm{d},q}$ and $t_\mathrm{m}$ as the detection time at the RX in the $q$th bit interval and the time of peak received signal, i.e., $t_\mathrm{m}=\displaystyle\arg \max_{t}\hat{P}(t)$, respectively. If $t_\mathrm{m}<T_\mathrm{b}$, we set $t_{\mathrm{d}, 1}=t_\mathrm{m}$. Otherwise, we set $t_{\mathrm{d}, 1}=T_\mathrm{b}$. Then, we have $t_{\mathrm{d},q}=t_{\mathrm{d},1}+(q-1)T_\mathrm{b}$. The number of molecules observed at $t_{\mathrm{d}, q}$, when molecules start being released at the beginning of the $g$th bit interval, $1\leq g\leq q$, is given by $N_\mathrm{v}\eta \hat{P}\left((q-g)T_\mathrm{b}+t_{\mathrm{d},1}\right)$. We denote $N_\mathrm{q}$ as the total number of molecules observed at the detection time in the $q$th bit interval. As we model the generation time of each vesicle within the TX as a continuous random process, the time interval between generating two successive vesicles is random. Thus, the generation time of the molecules encapsulated in different vesicles is different, which means that the received signal follows a Poisson binomial distribution since the molecules from different vesicles have a different probability of being observed at the detection time within the RX. According to Le Cam's theorem \cite{le1960approximation}, we approximate the Poisson binomial distribution by a Poisson distribution since the Poisson binomial distribution is cumbersome to work with. This approximation becomes accurate when the number of emitted molecules is large and the probability of molecules being observed within the RX, i.e., $\hat{P}(t)$, is small. Therefore, we model $N_q$ as RV that follows $N_q\sim\mathrm{Poiss}(\psi)$, where $\mathrm{Poiss}(\cdot)$ represents a Poisson distribution and $\psi$ is given by
	\begin{align}
		\psi=N_\mathrm{v}\eta\sum_{g=1}^{q}b_g\hat{P}\left((q-g)T_\mathrm{b}+t_{\mathrm{d}, 1}\right).
	\end{align}
	
	At the RX, we adopt a threshold detector that compares $N_q$ with a decision threshold to detect $b_q$. We model the threshold detector as 
	\begin{align}\label{h}
		\hat{b}_q=\left\{\begin{array}{lr}
			1, ~~\mathrm{if}\;N_q\geq\omega,
			\\
			0, ~~\mathrm{if}\;N_q<\omega,
		\end{array}
		\right.
	\end{align}
	where $\hat{b}_q$ is the detected bit corresponding to $b_q$ and $\omega$ is the decision threshold. We then denote $\Phi[q|\mathbf{b}_{1:q-1}]$ as the BER of the $q$th bit given the previously transmitted sequence $\mathbf{b}_{1:q-1}$. According to \eqref{h}, $\Phi[q|\mathbf{b}_{1:q-1}]$ is given by
	\begin{align}
		\Phi[q|\mathbf{b}_{1:q-1}]=P_1\mathrm{Pr}(N_q<\omega|N_q=1, \mathbf{b}_{1:q-1})+P_0\mathrm{Pr}(N_q\geq\omega|N_q=0, \mathbf{b}_{1:q-1}).
	\end{align}
	
	We further denote $\overline{\Phi}$ as the average BER over all realizations of $\mathbf{b}_{1:q}$ and all bits $b_q, q\in\{1, ..., Q\}$. Finally, we derive $\overline{\Phi}$ as
	\begin{align}
		\overline{\Phi}=\frac{1}{Q}\sum_{q=1}^{Q}\frac{\sum_{\mathbf{b}_{1:q-1}}\Phi[q|\mathbf{b}_{1:q-1}]}{2^{q-1}}.
	\end{align}
	\section{Numerical Results}\label{nr}
	\begin{table}[!t]
		\newcommand{\tabincell}[2]{\begin{tabular}{@{}#1@{}}#2\end{tabular}}
		\centering
		\caption{Simulation Parameters for Numerical Results}\label{tabl}
		\begin{tabular}{|c|c|c|c|}
			\hline
			\textbf{Parameter}&\textbf{Variable}&\textbf{Value}&\textbf{Reference}\\
			\hline
			Radius of the TX&$r_{\ss\T}$&$5\;\mu\mathrm{m}$&\cite{hat2011b}\\
			\hline
			Radius of the RX&$r_{\ss\R}$&$10\;\mu\mathrm{m}$&\cite{hat2011b}\\
			\hline
			Number of generated vesicles for bit 1&$N_\mathrm{v}$&200&\cite{huang2021membrane}\\
			\hline
			Number of molecules stored in each vesicle&$\eta$&20&\\
			\hline
			Generation rate of vesicles&$\mu$&$50, 100, 200\;\mathrm{s}^{-1}$&\\
			\hline
			Diffusion coefficient of vesicles&$D_\mathrm{v}$&$9\;\mu\mathrm{m}^2/\mathrm{s}$&\cite{kyoung2008vesicle}\\
			\hline
			Forward reaction rate constant of vesicles and the TX membrane&$k_\mathrm{f}$&$30\;\mu\mathrm{m}/\mathrm{s}$&\cite{huang2021membrane}\\
			\hline
			Ratio of the total area of receptors to the RX surface&$\mathcal{A}$&0.1&\cite{lindsay2017first}\\
			\hline
			Number of receptors&$N_\mathrm{r}$&1, 4, 11&\cite{li2004cellular}\\
			\hline
			Distance between the centers of the TX and RX&$r_0$&$20\;\mu\mathrm{m}$&\cite{yilmaz2014three}\\
			\hline
			Molecule diffusion coefficient&$D_\sigma$&$79.4\;\mu\mathrm{m}^2/\mathrm{s}$&\cite{yilmaz2014three}\\
			\hline
			Molecule degradation rate constant&$k_\mathrm{d}$&$0.8\;\mathrm{s}^{-1}$&\cite{deng2015modeling}\\
			\hline
			Length of binary sequence&$Q$&10&\\
			\hline
			Bit interval length&$T_\mathrm{b}$&$1.8\;\mathrm{s}$&\\
			\hline
			Probability of transmitting 0/1&$P_0/P_1$&0.5&\\
			\hline
		\end{tabular}
	\end{table}
	In this section, we present numerical results to validate our theoretical analysis and offer insightful discussions. Specifically, we use PBSs to simulate the random diffusion of molecules. In our simulations, we model the vesicle generation process as a 1D PPP, resulting in the time interval between two consecutive generated vesicles following an exponential distribution with a mean of $1/\mu$. After vesicles are generated, they perform random diffusion with a variance of $2D_\mathrm{v}\Delta t_\mathrm{s}$ and fuse to the TX membrane with a probability of $k_\mathrm{f}\sqrt{\frac{\pi\Delta t_\mathrm{s}}{D_\mathrm{v}}}$, where $\Delta t_\mathrm{s}$ is the simulation step. The detailed simulation framework for modeling the diffusion of vesicles within the TX and the MF process at the TX membrane is illustrated in \cite[Sec. VI]{huang2021membrane}. After the molecules have been released from the TX, we record their positions in each simulation step. If the position of a molecule at the end of the current simulation step is inside the TX volume, we assume that this molecule has hit the TX membrane in this simulation step. The coordinates of the hitting points on the TX membrane are calculated by using \cite[Eqs. (36)-(38)]{huang2021membrane}. If the coordinates of the hitting point of a molecule are inside a receptor, we treat this molecule as an absorbed molecule. Otherwise, the molecule is reflected back to the position it was at the start of the current simulation step \cite{ahmadzadeh2016comprehensive}. We choose the simulation time step as $\Delta t_\mathrm{s}=10^{-6}\;\mathrm{s}$ and all results are averaged over 1000 realizations. Throughout this section, we set the simulation parameters as shown in Table \ref{tabl}, unless otherwise stated. We note that NFM is only considered in Section \ref{ao}. In Figs. \ref{mrs}-\ref{ss}, Fig. \ref{thr}, and Fig. \ref{n}, we observe that the simulation results (denoted by markers) match well with the derived analytical curves (denoted by solid and dashed lines), which validates the accuracy of analytical derivation in Sections \ref{aor}, \ref{aors}, and \ref{nf}.
	\subsection{TX Model}
	In this subsection, we evaluate the molecule release rate and the number of harvested molecules at the TX.
	
	\begin{figure}[!t]
		\begin{center}
			\includegraphics[height=2.1in,width=0.42\columnwidth]{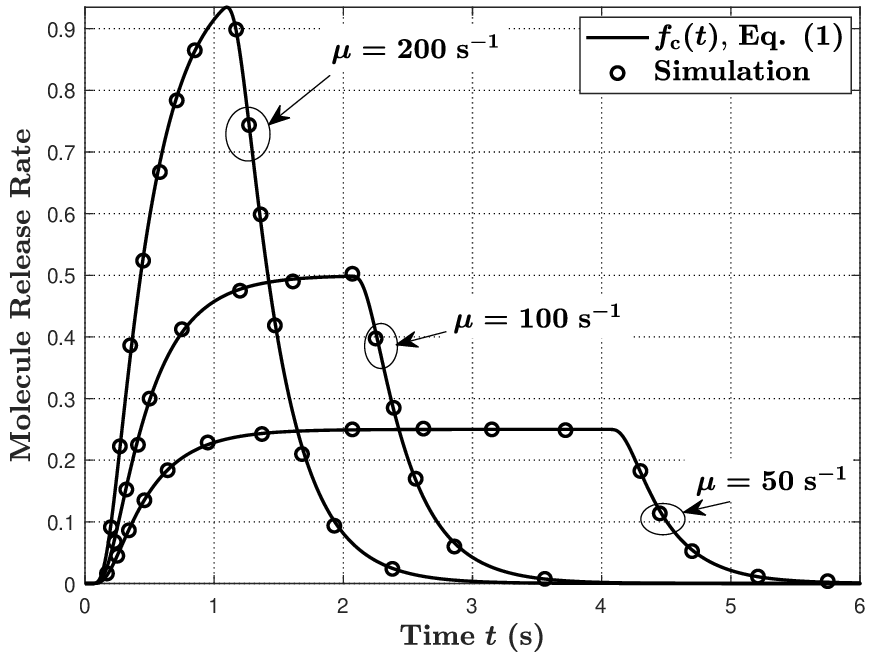}
			\caption{Molecule release rate from the TX versus time $t$ for different values of $\mu$.}\label{mrs}\vspace{-0.5em}
		\end{center}\vspace{-4mm}
	\end{figure}
	In Fig. \ref{mrs}, we plot the molecule release rate $f_\mathrm{c}(t)$ versus time $t$ for different values of $\mu$. First, we observe that a larger $\mu$ lead to a higher molecule release rate. This is because a faster generation of vesicles means that more molecules can be released from the TX within a small time interval. Second, when $\mu$ is small, $f_\mathrm{c}(t)$ maintains a constant value for a period of time. When $\mu$ is large, $f_\mathrm{c}(t)$ first increases and then decreases after reaching the maximum value. This is because small $\mu$ leads to a long emission period such that the concentration distribution of the molecules within the TX becomes stable, which results in a constant molecule release rate. 
	
	\begin{figure}[!t]
		\centering
		
		\subfigure[Number of absorbed molecules.]{
			\begin{minipage}[t]{0.5\linewidth}
				\centering
				\includegraphics[width=2.6in]{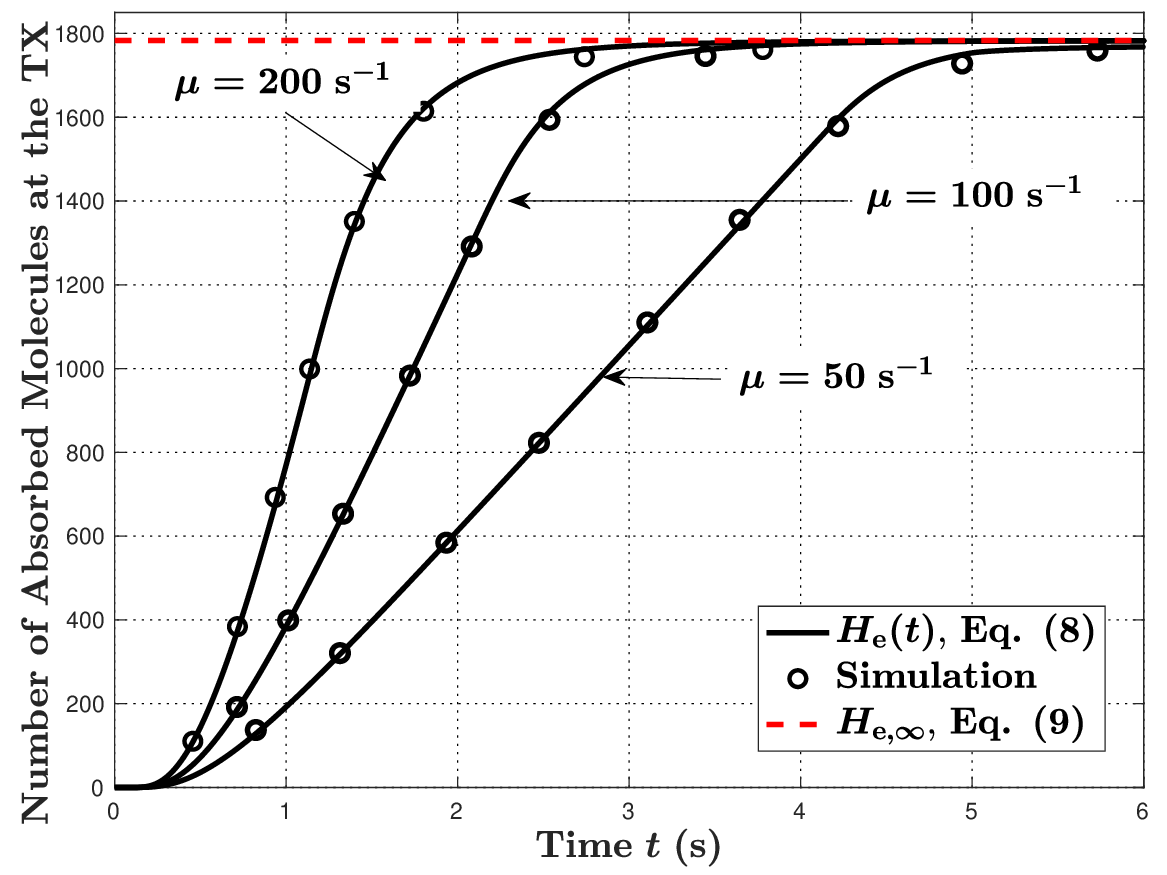}
				\label{a}
			\end{minipage}%
		}%
		\subfigure[Number of absorbed molecules for $\mu=200\;\mathrm{s}^{-1}$.]{
			\begin{minipage}[t]{0.5\linewidth}
				\centering
				\includegraphics[width=2.6in]{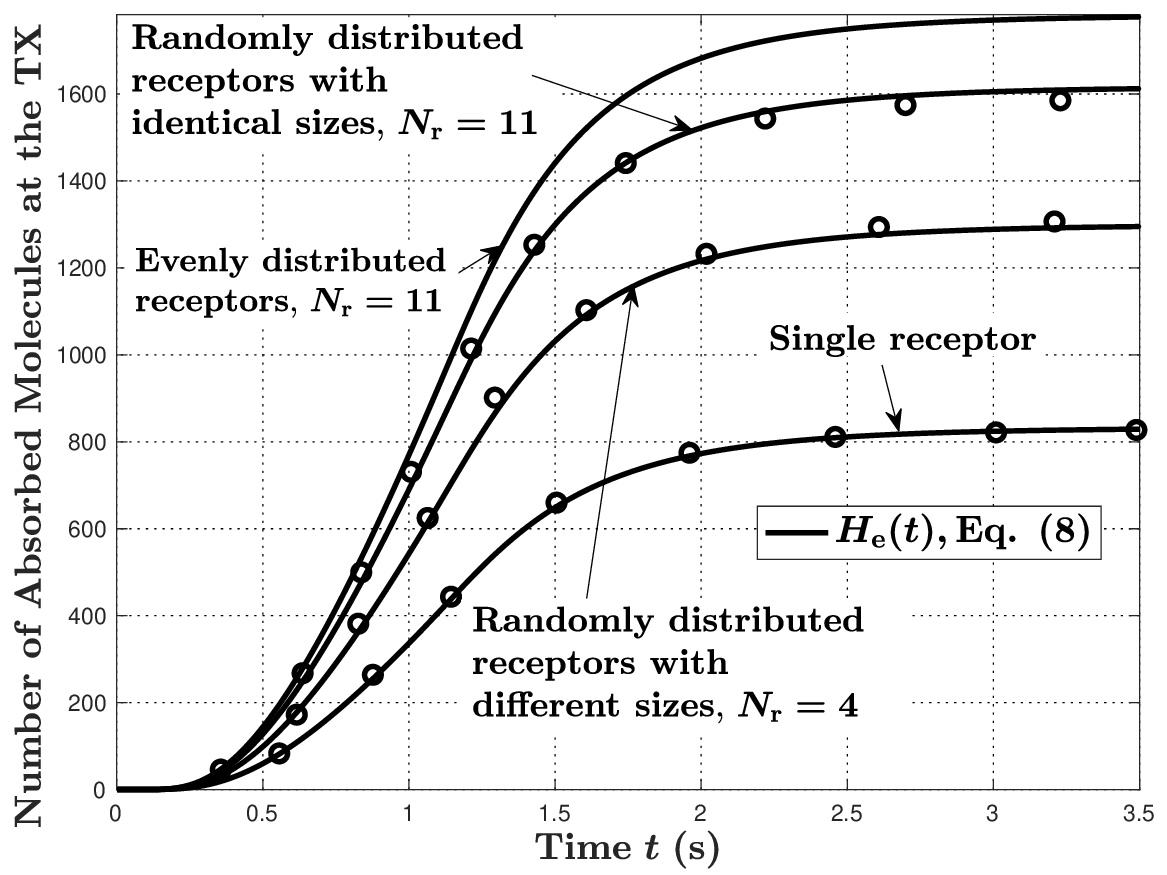}
				\label{b}
			\end{minipage}%
		}
		\quad
		\subfigure[Absorption rate of molecules]{
			\begin{minipage}[t]{0.5\linewidth}
				\centering
				\includegraphics[width=2.6in]{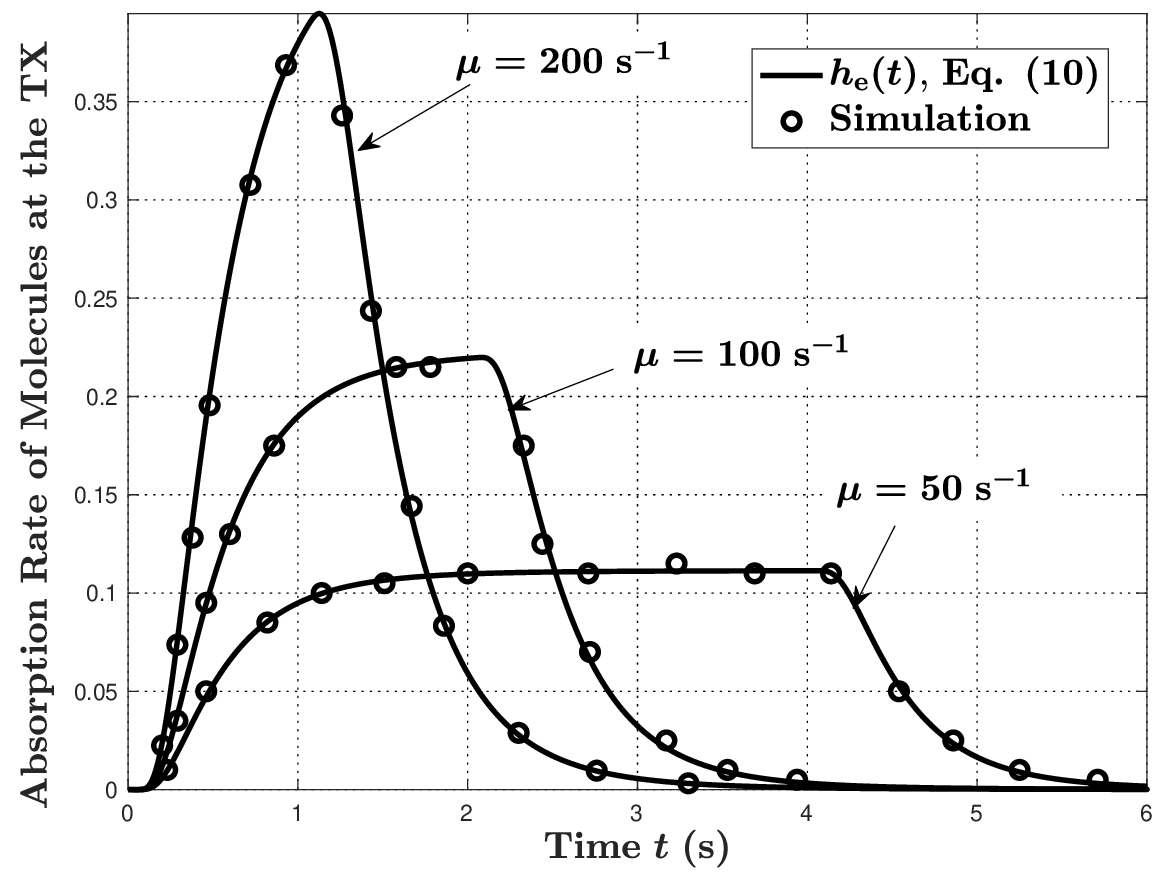}
				\label{c}
			\end{minipage}%
		}
		\centering
		\caption{(a): Number of molecules absorbed by the TX versus time $t$ for evenly distributed receptors and different $\mu$. (b): Number of  molecules absorbed by the TX versus time $t$ for different numbers, sizes, and distributions of receptors. (c): Absorption rate of molecules at the TX for evenly distributed receptors versus time $t$.}
		\label{ame}
	\end{figure}
	
	In Fig. \ref{ame}, we plot the number of molecules absorbed by the TX until time $t$ versus time $t$ in Fig. \ref{a} and Fig. \ref{b}, and the absorption rate of molecules at the TX versus time $t$ in Fig. \ref{c}. In Fig. \ref{a} and Fig. \ref{c}, we set $\mu\in\left\{50, 100, 200\right\}\;\mathrm{s}^{-1}$ and assume that receptors are evenly distributed over the TX membrane. Specifically, we apply the Fibonacci lattice \cite{gonzalez2010measurement} to determine the locations of the receptors that are evenly distributed. First, in Fig. \ref{a}, we observe that the total number of molecules that the TX can absorb is independent of the value of $\mu$, whch aligns with our derivation in \eqref{hei} and explanation in Remark \ref{fe}. Second, in Fig. \ref{b}, we compare the number of absorbed molecules for randomly distributed receptors, evenly distributed receptors, and a single receptor on the TX membrane. For receptors that are randomly distributed, we consider receptors that either have the same size or different sizes. For receptors with different sizes, we set their areas as $\mathcal{A}_1=0.01$, $\mathcal{A}_2=0.02$, $\mathcal{A}_3=0.03$, and $\mathcal{A}_4=0.04$, and their locations as $\vec{l}_1=[5\;\mu\mathrm{m}, \pi/2, \pi]$, $\vec{l}_2=[5\;\mu\mathrm{m}, \pi/2, \pi/2]$, $\vec{l}_3=[5\;\mu\mathrm{m}, \pi/2, 0]$, and $\vec{l}_4=[5\;\mu\mathrm{m}, \pi/2, 3\pi/2]$. For the single receptor, we set the location as $\vec{l}_\mathrm{s}=[-r_{\ss\T}, 0, 0]$. Moreover, we observe that the number of molecules absorbed by the evenly distributed receptors is larger than that by both the randomly distributed receptors and the single receptor. This is because evenly distributed receptors maintain an equal spacing, effectively covering the entire TX surface. Thus, the receptors have a higher probability of absorbing molecules. Third, in Fig. \ref{c}, we assume that receptors are evenly distributed over the TX membrane and observe that a larger $\mu$ leads to a higher peak molecule absorption rate. This is because the absorption rate of molecules is directly determined by the release rate of molecules, and a large $\mu$ leads to a higher peak molecule release rate.
	\subsection{Received Signals at the RX}
	\begin{figure}[!t]
		\centering
		
		\subfigure[Evenly distributed receptors, $N_\mathrm{r}=11$.]{
			\begin{minipage}[t]{0.5\linewidth}
				\centering
				\includegraphics[width=2.6in]{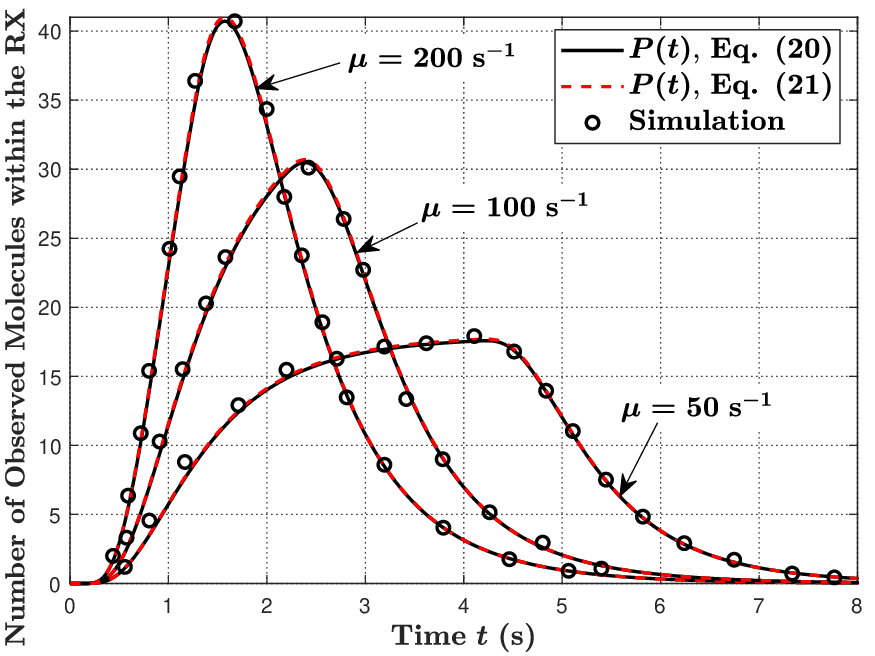}
				\label{v1}
			\end{minipage}%
		}%
		\subfigure[Other distributions of receptors, $\mu=200\;\mathrm{s}^{-1}$.]{
			\begin{minipage}[t]{0.5\linewidth}
				\centering
				\includegraphics[width=2.6in]{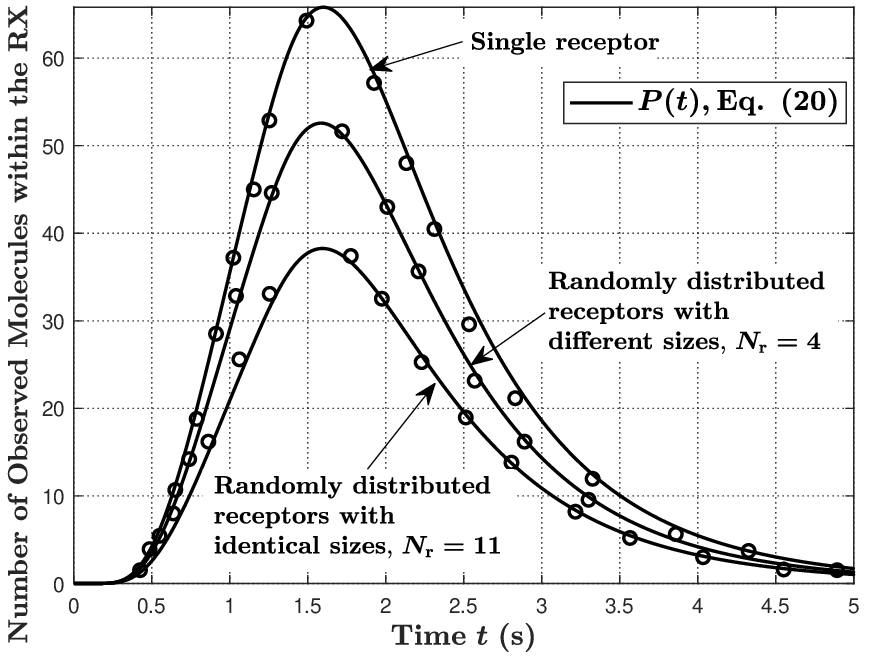}
				\label{v2}
			\end{minipage}%
		}
		\centering
		\caption{Number of observed molecules within the RX at time $t$ versus time $t$ for different distributions of receptors and different $\mu$.}
		\label{vv}
	\end{figure}
	In Fig. \ref{vv}, we plot the number of molecules observed within the RX $N_\mathrm{v}\eta P(t)$ versus time $t$, where the parameter setting for locations and sizes of receptors and the vesicle generation rate are the same as in Fig. \ref{ame}. First, in Fig. \ref{v1}, when receptors have identical sizes and are evenly distributed on the TX membrane, we observe that $P(t)$ in \eqref{pts} matches well with $P(t)$ in \eqref{ptf} and the simulation results, which demonstrates the accuracy of the simplification in \eqref{pts}. Second, in Fig. \ref{v1}, we observe that a larger $\mu$ leads to a higher peak received signal. This is because a larger $\mu$ leads to a faster release of molecules into the propagation environment such that more molecules can be observed within the RX at the same time. Third, Fig. \ref{v1} and Fig. \ref{v2} show that the received signal for the single-receptor TX is larger than that for the TX with evenly or randomly distributed receptors. This is because the single-receptor TX leads to fewer absorbed molecules than the other two distributions of receptors. Based on Fig. \ref{ame} and Fig. \ref{vv}, we observe that a larger number of absorbed molecules leads to a smaller received signal. This indicates a trade-off between energy efficiency and error probability, which is discussed more in detail in the next subsection.
	\subsection{Analysis of NFM and Error Performance}\label{ao}
	In this subsection, we integrate NFM into the TX model and examine its influence on the error probability and the number of absorbed molecules. For simplicity, we use a single-receptor TX as an example in this section, with $\vec{l}_\mathrm{s}=[-r_{\ss\T}, 0, 0]$ representing the location farthest  from the center of the RX, unless otherwise stated. 
	
		\begin{figure}[!t]
		\centering
		
		\subfigure[]{
			\begin{minipage}[t]{0.5\linewidth}
				\centering
				\includegraphics[width=2.6in]{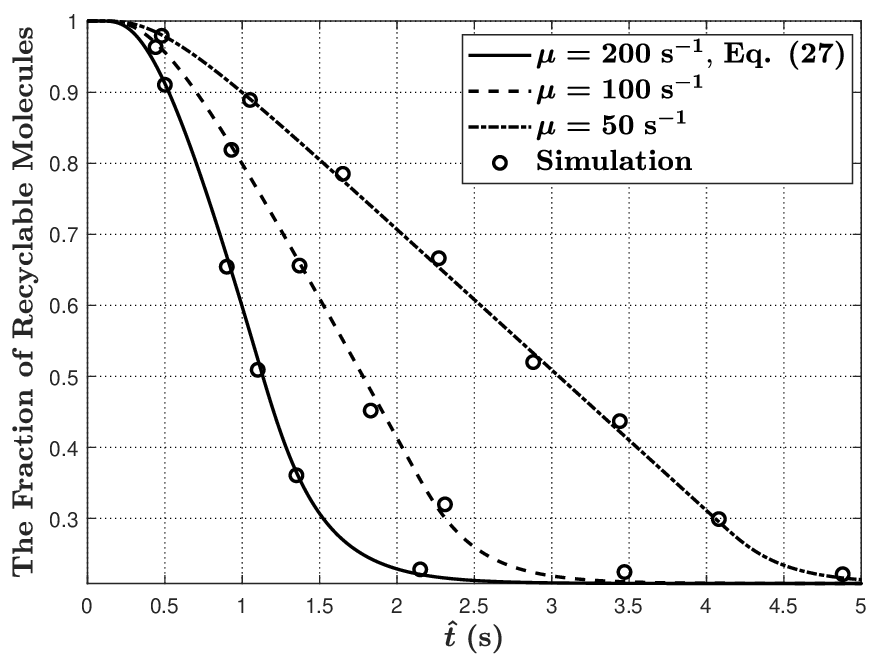}
				\label{s1}
			\end{minipage}%
		}%
		\subfigure[]{
			\begin{minipage}[t]{0.5\linewidth}
				\centering
				\includegraphics[width=2.6in]{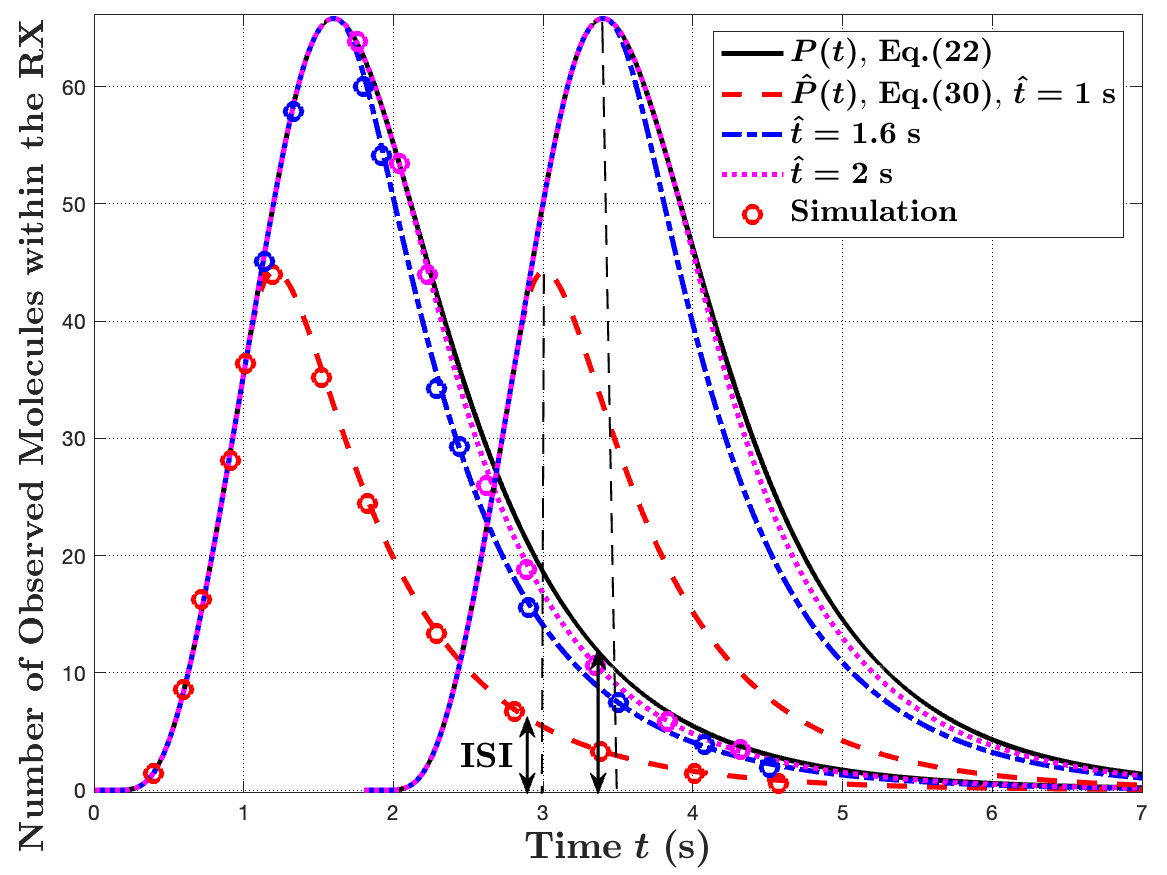}
				\label{s2}
			\end{minipage}%
		}
		\centering
		\caption{(a): Fraction of recyclable molecules at the TX versus $\hat{t}$ for different $\mu$. (b): Number of observed molecules within the RX versus time $t$ for different $\hat{t}$, where $\mu=200\;\mathrm{s}^{-1}$ and $T_\mathrm{b}=1.8\;\mathrm{s}$.}
		\label{ss}
	\end{figure}
\begin{figure}[!t]
	\centering
	
	\subfigure[Average BER without NFM]{
		\begin{minipage}[t]{0.5\linewidth}
			\centering
			\includegraphics[width=2.6in]{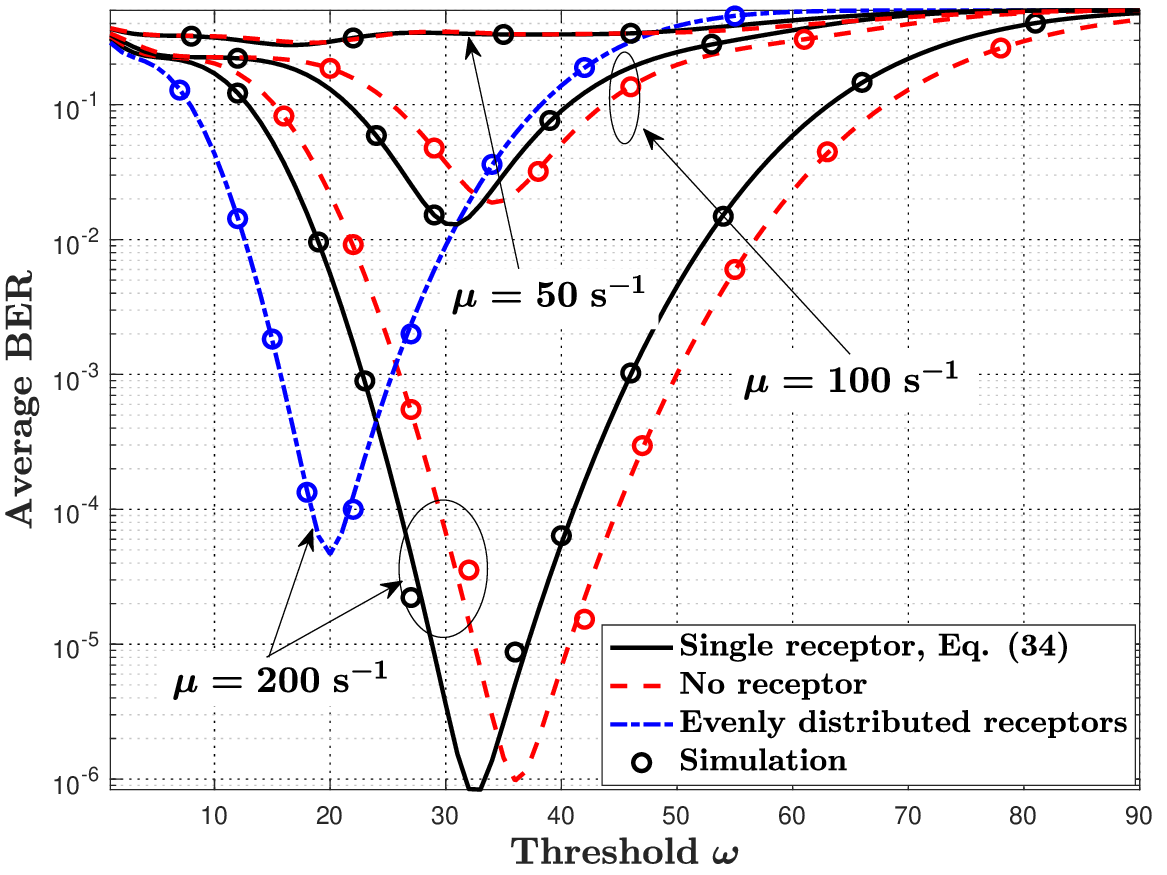}
			\label{thr}
		\end{minipage}%
	}%
	\subfigure[Average BER with NFM]{
		\begin{minipage}[t]{0.5\linewidth}
			\centering
			\includegraphics[width=2.6in]{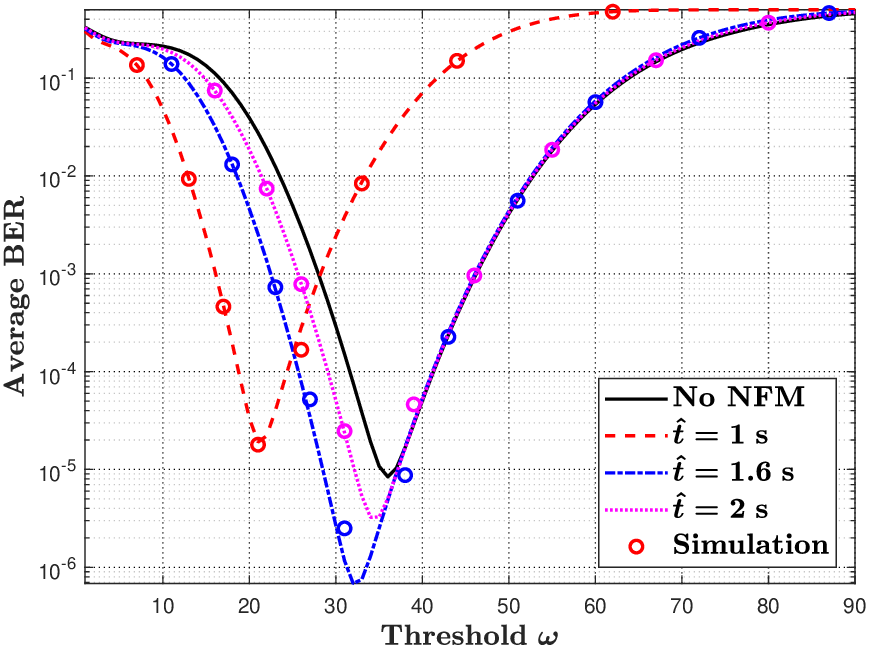}
			\label{n}
		\end{minipage}%
	}
	\quad
	\subfigure[Average minimum BER with NFM]{
		\begin{minipage}[t]{0.5\linewidth}
			\centering
			\includegraphics[width=2.6in]{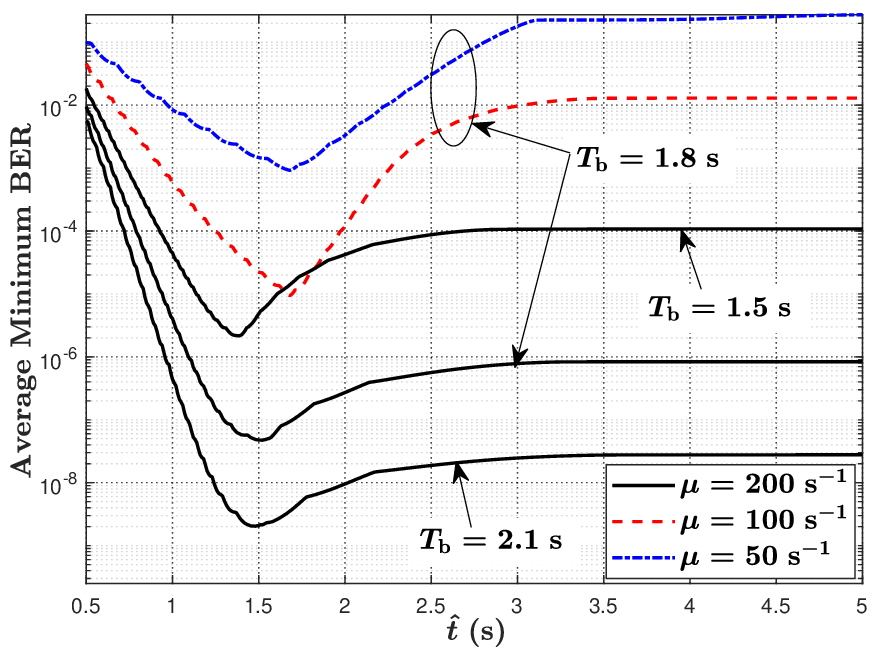}
			\label{am}
		\end{minipage}%
	}
	\centering
	\caption{(a): Average BER versus threshold $\omega$ for different $\mu$. (b): Average BER versus threshold $\omega$ for different $\hat{t}$. (c): Average minimum BER versus $\hat{t}$ for different $\mu$ and $T_\mathrm{b}$.}
	\label{tn}
\end{figure}

In Fig. \ref{ss}, we assume the TX stops releasing molecules at time $\hat{t}$ due to NFM and plot the fraction of molecules that can be reused by the TX versus $\hat{t}$ for different $\mu$ in Fig. \ref{s1} and the received signal at the RX versus time $t$ for different $\hat{t}$ in Fig. \ref{s2}. First, in Fig. \ref{s1}, we observe that the fraction of recyclable molecules decreases as $\hat{t}$ increases. This is due to the fact that the TX releases molecules for a longer time when $\hat{t}$ is larger such that fewer molecules are left for the TX to reuse. Second, as larger $\mu$ leads to a faster release of molecules, the fraction of recyclable molecules decreases as $\mu$ increases, as shown in Fig. \ref{s1}. Third, in Fig. \ref{s2}, we plot multiple pulses of the received signal at the RX in \eqref{ptf} when NFM is not employed and the received signal in \eqref{tp} for NFM, where $T_\mathrm{b}=1.8\;\mathrm{s}$. In this figure, we set $\hat{t}=\left\{1, 1.6, 2\right\}\;\mathrm{s}$, which corresponds to the TX stopping releasing molecules when the received signal is increasing, reaches the maximum value, and is decreasing, respectively. As we adopt a threshold-based detector as in \eqref{h}, the error probability of detection at the RX is mainly influenced by the peak received signal and ISI. In this figure, signal detection is performed at the maximum value, and the ISI caused by the previous symbol is highlighted. When the TX stops releasing molecules at $\hat{t}=1\;\mathrm{s}$, we observe that the received signal has less ISI, but also has a smaller peak value. When the TX stops releasing molecules at $\hat{t}=1.6\;\mathrm{s}$, we observe that the received signal reaches the peak value of $P(t)$ and still causes less ISI. When the TX stops releasing molecules at $\hat{t}=2\;\mathrm{s}$, the received signal can achieve the peak value, but causes more ISI compared to the scenario when $\hat{t}=1.6\;\mathrm{s}$. Therefore, choosing an appropriate $\hat{t}$ is important for maximum performance. Notably, even though the differences in ISIs between $\hat{t}=1.6\;\mathrm{s}$, $\hat{t}=2\;\mathrm{s}$, and not incorporating NFM are small, they lead to distinct average BERs as shown in Fig. \ref{n}. 

	\begin{table}[!t]
	\newcommand{\tabincell}[2]{\begin{tabular}{@{}#1@{}}#2\end{tabular}}
	\centering
	\caption{Minimum BER and Fraction of Recyclable Molecules for Different $\mu$ and $T_\mathrm{b}$}\label{tab3}
	\begin{tabular}{|c|c|c|c|c|c|}
		\hline
		$\mu\;[\mathrm{s}^{-1}]$&Optimal $\hat{t}\;[\mathrm{s}]$&Min. BER&Recyclable Molecules (with NFM)&Recyclable Molecules (without NFM)&$T_\mathrm{b}\;[\mathrm{s}]$\\
		\hline
		$50$&$1.68$&$9.21\times10^{-4}$&$76.98\;\%$&$4.89\;\%$&$1.8$\\
		\hline
		$100$&$1.68$&$9.4\times10^{-6}$&$53.93\;\%$&$4.89\;\%$&$1.8$\\
		\hline
		$200$&$1.52$&$4.73\times10^{-8}$&$30.09\;\%$&$4.89\;\%$&$1.8$\\
		\hline
		$200$&$1.38$&$2.17\times10^{-6}$&$35.13\;\%$&$4.89\;\%$&$1.5$\\
		\hline
		$200$&$1.48$&$2.09\times10^{-9}$&$31.32\;\%$&$4.89\;\%$&$2.1$\\
		\hline
	\end{tabular}
\end{table}

 In Fig. \ref{tn}, we examine the effects of $\mu$, $\hat{t}$, and $T_\mathrm{b}$ on the average BER. Due to the long PBS times, we employ Monte Carlo simulation in this figure, using the Poisson approximation to generate the number of observed molecules. We note that the derived analytical result for the CIR has been verified by PBSs, and the Poisson approximation for the number of observed molecules has been widely adopted in previous studies \cite{jamali2019channel}.  First, in Fig. \ref{thr}, we compare TXs with a single receptor, evenly distributed receptors, and no receptor, respectively, plotting the BER for $\mu\in\left\{50, 100, 200\right\}\;\mathrm{s}^{-1}$ if NFM is not employed. We observe that the BER for the TX with evenly distributed receptors is significantly larger than that of the TXs with a single receptor and no receptors, respectively. This is because the TX with evenly distributed receptors absorbs more molecules, resulting in a lower received signal at the RX. We also note that the BER increases as $\mu$ decreases, since a larger $\mu$ leads to a higher received signal as illustrated in Fig. \ref{v1}. Second, in Fig. \ref{n}, we incorporate NFM at the TX and assume that the TX stops releasing molecules after $\hat{t}$ for each emission. We observe that the TX with NFM can achieve a lower BER when $\hat{t}\in\left\{1.6, 2\right\}\mathrm{s}$, compared to the scenario without NFM or without considering molecule harvesting. This occurs because the received signal can attain the same peak value as in the scenario without NFM, while also causing less ISI as shown in Fig. \ref{s2}. Third, from Fig. \ref{thr} and Fig. \ref{n}, we observe that BER can be minimized by choosing an appropriate threshold. In Fig. \ref{am}, we plot the minimum BER versus $\hat{t}$ for different $T_\mathrm{b}$ and $\mu$, and observe that the minimum BER decreases as $T_\mathrm{b}$ or $\mu$ increase. We also list the optimal $\hat{t}$ achieving the lowest BER in Table \ref{tab3} and the corresponding fraction of recyclable molecules with NFM and without NFM, respectively. Based on Fig. \ref{tn} and Table \ref{tab3}, we clearly observe the \textit{significant advantages} of incorporating NFM at the TX, including the improvement of the system performance and energy efficiency. On the one hand, the system with NFM can achieve a considerably lower BER, e.g., $4.73\times10^{-8}$, compared to the system without NFM or without employing molecule harvesting. On the other hand, the system with NFM can recycle up to $30.09\%$ of the molecules for subsequent emission rounds, contributing to a higher energy efficiency than the system without NFM. Fourth, in Table \ref{tab3}, we observe a trade-off between error performance and energy efficiency, as higher minimum BERs are associated with larger numbers of absorbed molecules. This balance is determined by parameter $\hat{t}$. Consequently, the system can be tailored to meet specific application requirements by adjusting this parameter, striking an appropriate balance between error performance and energy efficiency.
 
 \section{Conclusion}\label{con}
 In this paper, we examined a molecule harvesting TX model whose membrane is covered by heterogeneous receptors of varying sizes and arbitrary locations. By considering continuous vesicle generation within the TX and a transparent RX, we derived the molecule release rate, the number of absorbed molecules at the TX, and the received signal at the RX. We further integrated NFM at the TX and calculated the fraction of molecules recyclable for subsequent emissions. PBSs results verified the accuracy of the derived expressions. Moreover, numerical results revealed that a TX with evenly distributed receptors can absorb more molecules than a TX with randomly distributed receptors or a single large receptor. Our findings also highlighted the advantages of NFM in mitigating ISI and enhancing energy efficiency compared to a TX that only employs molecule harvesting or a TX that disregards both NFM and molecule harvesting. In addition, our results illustrated a trade-off between error performance and energy efficiency in the considered MC system.
 
This paper focused mainly on the absorption of molecules, while the subsequent processes within the TX related to recycling molecules, such as breaking down or repackaging molecules into vesicles, were neglected. The impact of these processes on other performance indicators of the MC system, such as latency, capacity, and error performance, especially in resource-constrained environments, is an interesting topic for further investigation.

	\appendices
	\section{Proof of Theorem \ref{t1}}\label{A1}
	We first consider $0<t\leq\tau$. When a vesicle is generated in the center of the TX at time $u$, $0\leq u\leq\tau$, the probability that this vesicle fuses with the TX membrane at time $t$ is given by $f_\mathrm{r}(t-u)$. Here, the molecule release rate equals the vesicle fusion probability since MF guarantees the release of molecules into the propagation environment. Due to the continuous generation of vesicles, the number of vesicles fusing with the TX membrane during time interval $[t, t+\delta t]$ is given by $\mu\int_{0}^{t}f_\mathrm{r}(t-u)\;\mathrm{d}u$. As we define $f_\mathrm{c}(t)$ as the probability of a vesicle fusing with the TX membrane during an infinitesimally small time interval, we obtain $f_\mathrm{c}(t)$ as
	\begin{align}\label{fcp}
		f_\mathrm{c}(t)=\frac{\mu}{N_\mathrm{v}}\int_{0}^{t}f_\mathrm{r}(t-u)\;\mathrm{d}u=\frac{\mu}{N_\mathrm{v}}\int_{0}^{t}f_\mathrm{r}(u)\;\mathrm{d}u.
	\end{align}
	By substituting \cite[Eq. (5)]{huang2021membrane} into \eqref{fcp}, we obtain \eqref{fc}. 
	Second, we consider $t>\tau$. With this condition, all considered vesicles have already been generated by the TX. Therefore, the number of vesicles fusing with the TX membrane during time interval $[t, t+\delta t]$ is given by $\mu\int_{0}^{\tau}f_\mathrm{r}(t-u)\;\mathrm{d}u$. Then, we obtain $f_\mathrm{c}(t)$ as
	\begin{align}\label{fcp2}
		f_\mathrm{c}(t)=\frac{\mu}{N_\mathrm{v}}\int_{0}^{\tau}f_\mathrm{r}(t-u)\;\mathrm{d}u=\frac{\mu}{N_\mathrm{v}}\int_{t-\tau}^{t}f_\mathrm{r}(u)\;\mathrm{d}u.
	\end{align} 
	By substituting \cite[Eq. (5)]{huang2021membrane} into \eqref{fcp2}, we obtain \eqref{fc2}.
	
	\section{Proof of Corollary \ref{c1}}\label{A7}
	According to \eqref{bk}, we first express $f_\mathrm{c}(t)$ as 
	\begin{align}
		f_\mathrm{c}(t)=f_{\mathrm{c}, 1}(t)u(t)-\left(f_{\mathrm{c}, 1}(t)-f_{\mathrm{c}, 2}(t)\right)u(t-\tau),
	\end{align}
	where $u(t)$ is the unit step function. To derive $\lim\limits_{t\rightarrow\infty}H_\mathrm{e}(t)$, we perform Laplace transform on \eqref{ge}, which yields $\mathcal{H}_\mathrm{e}(s)=\mathcal{F}_\mathrm{c}(s)\mathcal{H}(s)$, where $\mathcal{H}_\mathrm{e}(s)$, $\mathcal{F}_\mathrm{c}(s)$, and $\mathcal{H}(s)$ are the Laplace transform of $H_\mathrm{e}(t)$, $f_\mathrm{c}(t)$, and $H(t)$, respectively. According to \cite[Eqs. (1.8), (2.2), (5.1), (17.62), (17.63)]{oberhettinger2012tables}, we derive $\mathcal{F}_\mathrm{c}(s)$ and $\mathcal{H}(s)$ as
	\begin{align}\label{t}
		\mathcal{F}_\mathrm{c}(s)=\frac{4r_{\ss\T}^2k_\mathrm{f}\mu\left(1-\exp(-\tau s)\right)}{N_\mathrm{v}s}\sum_{n=1}^{\infty}\frac{\lambda_n^3j_0(\lambda_nr_{\ss\T})}{\left(2\lambda_nr_{\ss\T}-\sin(2\lambda_nr_{\ss\T})\right)\left(D_\mathrm{v}\lambda_n^2+s\right)},
	\end{align}
	and 
	\begin{align}\label{mhs}
		\mathcal{H}(s)=\frac{w}{s\sqrt{D_\sigma(k_\mathrm{d}+s)}}-\frac{(\sqrt{s+k_\mathrm{d}}-\gamma\sqrt{D_\sigma})w\gamma}{\sqrt{s+k_\mathrm{d}}(s-\zeta)\zeta}-\frac{w\gamma^2}{\zeta s}\sqrt{\frac{D_\sigma}{k_\mathrm{d}+s}}+\frac{w\gamma}{\zeta s}.
	\end{align}	
	According to the final value theorem \cite[Eq. (1)]{chen2007final}, we have $\lim\limits_{t\rightarrow\infty}H_\mathrm{e}(t)=\lim\limits_{s\rightarrow 0}s\mathcal{F}(s)\mathcal{H}(s)$. By substituting \eqref{t} and \eqref{mhs} into this expression, we obtain 
	\begin{align}\label{ll}
		\lim\limits_{t\rightarrow\infty}H_\mathrm{e}(t)=\left(\frac{w}{\sqrt{D_\sigma k_\mathrm{d}}}-\frac{w\gamma^2}{\zeta}\sqrt{\frac{D_\sigma}{k_\mathrm{d}}}+\frac{w\gamma}{\zeta}\right)\frac{4r_{\ss\T}^2k_\mathrm{f}}{D_\mathrm{v}}\sum_{n=1}^{\infty}\frac{\lambda_nj_0(\lambda_nr_{\ss\T})}{2\lambda_nr_{\ss\T}-\sin(2\lambda_nr_{\ss\T})}.
	\end{align}
As the total fraction of released molecules approaches 1 when $t\rightarrow\infty$, according to \eqref{bk}, we have $\int_{0}^{\tau}f_{\mathrm{c}, 1}(t)\mathrm{d}t+\int_{\tau}^{\infty}f_{\mathrm{c}, 2}(t)\mathrm{d}t=1$. By substituting \eqref{fc} and \eqref{fc2} into this expression, we obtain $\sum_{n=1}^{\infty}\frac{\lambda_nj_0(\lambda_nr_{\ss\T})}{2\lambda_nr_{\ss\T}-\sin(2\lambda_nr_{\ss\T})}=\frac{D_\mathrm{v}}{4r_{\ss\T}^2k_\mathrm{f}}$. By substituting this expression into \eqref{ll}, we finally obtain \eqref{hei}.
	
	\section{Proof of Lemma \ref{l2}}\label{A3}
	\begin{figure}[!t]
		\begin{center}
			\includegraphics[height=1.5in,width=0.23\columnwidth]{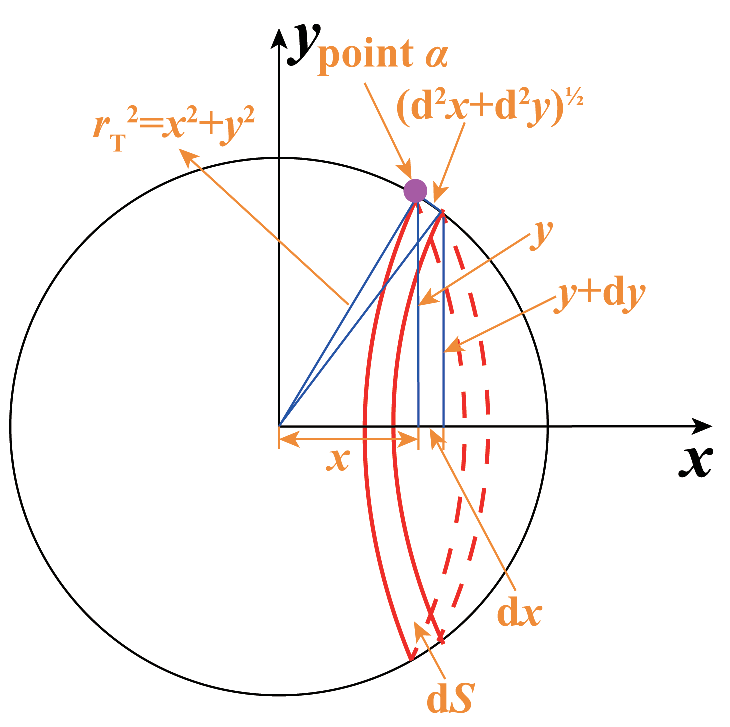}
		\caption{Mathematical model for the surface integral over the spherical $\mathrm{TX}$ membrane.}\label{integral}\vspace{-0.5em}
		\end{center}
		\vspace{-4mm}
	\end{figure}	
	We perform the surface integral by considering a small surface element that is the lateral surface of a conical frustum, denoted by $\mathrm{d}S$, on the TX membrane, as shown in Fig. \ref{integral}. The point $\alpha$ is on $\mathrm{d}S$ with the coordinates $(x, y, z)$. According to \cite[Appendix B]{huang2021membrane}, we obtain $\mathrm{d}S=2\pi r_{\ss\T}\mathrm{d}x$. As $\mathrm{d}S$ is infinitesimal, the distance between each point on $\mathrm{d}S$ and the center of the RX is given by $r_{\alpha}=\sqrt{r_{\ss\T}^2+r_0^2-2r_0x}$. Then, we perform the surface integral of $P_\alpha(t)$ over the TX membrane as
	\begin{align}\label{put}
		P_\mathrm{u}(t)=\frac{1}{2r_{\ss\T}}\int_{-r_{\ss\T}}^{r_{\ss\T}}P_\alpha(t)\big|_{r_\alpha=\sqrt{r_{\ss\T}^2+r_0^2-2r_0x}}\;\mathrm{d}x.
	\end{align}
	By substituting \eqref{pit} into \eqref{put}, we obtain \eqref{pu}.
	
	\section{Proof of Lemma \ref{l4}}\label{A4}
	We denote $P_{\mathrm{r},i}(t)$ as the probability that a molecule is observed at the RX at time $t$, under the assumption that molecules were continuously released from the $i$th receptor. Similar to the proof of Theorem \ref{t2}, as $h_{\mathrm{e},i}(u)$ is the molecule release rate from the $i$th receptor at time $u$, and $P_\alpha(t-u)\big|_{r_\alpha=d_i}$ is the probability that a molecule is observed at the RX at time $t$ assuming that this molecule was released from the $i$th receptor at time $u$, we obtain $P_{\mathrm{r},i}(t)$ as
	\begin{align}\label{p}
		P_{\mathrm{r},i}(t)=\int_{0}^{t}h_{\mathrm{e},i}(u)P_\alpha(t-u)\big|_{r_\alpha=d_i}\mathrm{d}u=h_{\mathrm{e},i}(t)*P_\alpha(t)\big|_{r_\alpha=d_i}.
	\end{align}
	Then, the probability that a molecule is observed at the RX assuming that molecules were released from all receptors is given by $P_\mathrm{r}(t)=\sum_{i=1}^{N_\mathrm{r}}P_{\mathrm{r},i}(t).$  By substituting \eqref{pit} and \eqref{heit} into \eqref{p} and this expression, we obtain \eqref{pr}.
	
	\section{Proof of Corollary \ref{c6}}\label{A5}
	When identical receptors are evenly distributed over the TX membrane, the membrane can be regarded as a homogeneous boundary such that the molecule release rate is the same at any point on the TX membrane. Therefore, given the molecule release rate $f_\mathrm{c}(t)$ and molecule absorption rate $h_\mathrm{e}(t)$ at the TX membrane, the net molecule release rate from the TX membrane is given by $f_\mathrm{c}(t)-h_\mathrm{e}(t)$. Then, $P(t)$ can be obtained as $P(t)=(f_\mathrm{c}(t)-h_\mathrm{e}(t))*P_\mathrm{u}(t)$. By substituting \eqref{he} into this expression, we obtain \eqref{pts}.
	
	When there is a single receptor on the TX membrane, we set $\mathcal{A}_i=\mathcal{A}$, $N_\mathrm{r}=1$, and $d_\mathrm{i}=d_\mathrm{s}$ in \eqref{ptf}, which leads to \eqref{ptfc}.
	
	\section{Proof of Theorem \ref{t4}}\label{A6}
	We first consider the scenario when $\hat{t}<\tau$. The derivation of $\beta_1(t\rightarrow\infty)$ is similar to that of $H_{\mathrm{e}, \infty}$ in Appendix \ref{A7}. By performing the Laplace transform of $\beta_1(t)=\hat{f}_{\mathrm{c}, 1}(t)*H(t)$, we obtain $\hat{\beta}_1(s)=\hat{\mathcal{F}}_{\mathrm{c}, 1}(s)\mathcal{H}(s)$, where $\hat{\beta}_1(s)$ and $\hat{\mathcal{F}}_{\mathrm{c}, 1}(s)$ are the Laplace transforms of $\beta_1(t)$ and $\hat{f}_{\mathrm{c}, 1}(t)$, respectively. According to \eqref{tbk}, we rewrite $\hat{f}_{\mathrm{c}, 1}(t)$ as 
	\begin{align}
		\hat{f}_{\mathrm{c}, 1}(t)=f_{\mathrm{c}, 1}(t)\left(u(t)-u(t-\hat{t})\right).
	\end{align}
According to \cite[Eqs. (1.8), (2.2)]{oberhettinger2012tables}, we obtain $\hat{\mathcal{F}}_{\mathrm{c}, 1}(s)$ as
\begin{align}\label{hmf}
	\hat{\mathcal{F}}_{\mathrm{c}, 1}(s)=\frac{4r_{\ss\T}^2 k_\mathrm{f}\mu}{N_\mathrm{v}D_\mathrm{v}}\sum_{n=1}^{\infty}\frac{\lambda_nj_0(\lambda_nr_{\ss\T})}{2\lambda_nr_{\ss\T}-\mathrm{sin}(2\lambda_nr_{\ss\T})}\left(\hat{t}-\frac{1-\exp(-D_\mathrm{v}\lambda_n^2\hat{t}}{D_\mathrm{v}\lambda_n^2}\right).
\end{align}
Based on the final value theorem \cite[Eq. (1)]{chen2007final}, we have $\beta_1(t\rightarrow\infty)=\lim\limits_{s\rightarrow 0}s\hat{\mathcal{F}}_{\mathrm{c}, 1}(s)\mathcal{H}(s)$. By substituting \eqref{hmf} and \eqref{mhs} into this expression, we obtain $\beta_1(t\rightarrow\infty)$ as
\begin{align}\label{l}
	\beta_1(t\!\rightarrow\!\infty\!)\!=\!\!\frac{4r_{\ss\T}^2 k_\mathrm{f}\mu}{N_\mathrm{v}D_\mathrm{v}}\!\!\left(\!\frac{w}{\sqrt{D_\sigma k_\mathrm{d}}}\!-\!\frac{w\gamma^2}{\zeta}\sqrt{\frac{D_\sigma}{k_\mathrm{d}}}+\frac{w\gamma}{\zeta}\right)\!\sum_{n=1}^{\infty}\frac{\lambda_nj_0(\lambda_nr_{\ss\T})}{2\lambda_nr_{\ss\T}-\mathrm{sin}(2\lambda_nr_{\ss\T})}\!\left(\!\hat{t}\!-\frac{1-\exp(-D_\mathrm{v}\lambda_n^2\hat{t})}{D_\mathrm{v}\lambda_n^2}\right).
\end{align}
We then derive $\beta_2(\hat{t})$. We note that $\beta_2(\hat{t})$ represents the fraction of molecules released after time $\hat{t}$. Therefore, we derive $\beta_2(\hat{t})$ as
\begin{align}\label{chi2}
	\beta_2(\hat{t})&=\int_{\hat{t}}^{\tau}f_{\mathrm{c}, 1}(t)\mathrm{d}t+\int_{\tau}^{\infty}f_{\mathrm{c}, 2}(t)\mathrm{d}t\notag\\&=\frac{4r_{\ss\T}^2 k_\mathrm{f}\mu}{N_\mathrm{v}D_\mathrm{v}}\sum_{n=1}^{\infty}\frac{\lambda_nj_0(\lambda_nr_{\ss\T})}{2\lambda_nr_{\ss\T}-\mathrm{sin}(2\lambda_nr_{\ss\T})}\left(\tau-\hat{t}+\frac{1-\exp(-D_\mathrm{v}\lambda_n^2\hat{t})}{D_\mathrm{v}\lambda_n^2}\right).
\end{align} 
By substituting \eqref{l} and \eqref{chi2} into \eqref{ch}, we obtain \eqref{ha}.

Next, we consider $\hat{t}>\tau$. The derivation process for $\chi_2(\hat{t})$ closely follow that for $\chi_1(\hat{t})$. We denote $\hat{\mathcal{F}}_{\mathrm{c}, 2}$ as the Laplace transform of $\hat{f}_{\mathrm{c}, 2}(t)$. According to \eqref{hf}, we re-write $\hat{f}_{\mathrm{c}, 2}(t)$ as
	\begin{align}
		\hat{f}_{\mathrm{c}, 2}(t)=f_{\mathrm{c},1}(t)u(t)-(f_{\mathrm{c},1}(t)-f_{\mathrm{c},2}(t))u(t-\tau)-f_{\mathrm{c},2}(t)u(t-\hat{t}).
	\end{align}
	According to \cite[Eqs. (1.8), (2.2), (5.1)]{oberhettinger2012tables}, we derive $\hat{\mathcal{F}}_{\mathrm{c}, 2}(s)$ as
	\begin{align}\label{tfc}
		\hat{\mathcal{F}}_{\mathrm{c}, 2}(s)=&\frac{4r_{\ss\T}^2 k_\mathrm{f}\mu}{N_\mathrm{v}D_\mathrm{v}}\sum_{n=1}^{\infty}\frac{\lambda_nj_0(\lambda_nr_{\ss\T})}{2\lambda_nr_{\ss\T}-\mathrm{sin}(2\lambda_nr_{\ss\T})}\left[\frac{D_\mathrm{v}\lambda_n^2}{s(D_\mathrm{v}\lambda_n^2+s)}\left(1-\exp(-\tau s)\right)\right.\notag\\&\left.-\frac{\exp\left(-D_\mathrm{v}\lambda_n^2(\hat{t}-\tau)\right)-\exp\left(-D_\mathrm{v}\lambda_n^2\hat{t}\right)}{s+D_\mathrm{v}\lambda_n^2}\exp(-\hat{t}s)\right].
	\end{align}
 By substituting \eqref{tfc} and \eqref{mhs} into $\beta_1(t\rightarrow\infty)=\lim\limits_{s\rightarrow 0}s\hat{\mathcal{F}}_{\mathrm{c}, 2}(s)\mathcal{H}(s)$, we obtain $\beta_1(t\rightarrow\infty)$ as 
	\begin{align}\label{th}
		\beta_1(t\rightarrow\infty)=&\frac{4r_{\ss\T}^2k_\mathrm{f}\mu}{N_\mathrm{v}D_\mathrm{v}}\left(\frac{w}{\sqrt{D_\sigma k_\mathrm{d}}}-\frac{w\gamma^2}{\zeta}\sqrt{\frac{D_\sigma}{k_\mathrm{d}}}+\frac{w\gamma}{\zeta}\right)\sum_{n=1}^{\infty}\frac{\lambda_nj_0(\lambda_nr_{\ss\T})}{2\lambda_nr_{\ss\T}-\sin(2\lambda_nr_{\ss\T})}\notag\\&\times\left(\tau-\frac{\exp\left(-D_\mathrm{v}\lambda_n^2(\hat{t}-\tau)\right)-\exp\left(-D_\mathrm{v}\lambda_n^2\hat{t}\right)}{D_\mathrm{v}\lambda_n^2}\right).
	\end{align}
We then obtain $\beta_2(\hat{t})$ as
	\begin{align}\label{f}
		\beta_2(\hat{t})&=\int_{\hat{t}}^{\infty}f_{\mathrm{c},2}(t)\mathrm{d}t\notag\\&=\frac{4r_{\ss\T}^2 k_\mathrm{f}\mu}{N_\mathrm{v}D_\mathrm{v}^2}\sum_{n=1}^{\infty}\frac{j_0(\lambda_nr_{\ss\T})}{(2\lambda_nr_{\ss\T}-\mathrm{sin}(2\lambda_nr_{\ss\T}))\lambda_n}\left[\exp\left(-D_\mathrm{v}\lambda_n^2(\hat{t}-\tau)\right)-\exp\left(-D_\mathrm{v}\lambda_n^2\hat{t}\right)\right].
	\end{align}
	By substituting \eqref{th} and \eqref{f} into \eqref{ch}, we get \eqref{chi}.
	
	\bibliographystyle{IEEEtran}
	\bibliography{ref}
\end{document}